\documentclass[letterpaper,11pt]{extarticle}

\usepackage[utf8]{inputenc}

\usepackage[american]{babel}

\usepackage{kotex}

\usepackage{subfiles}

\usepackage{xr-hyper}

\makeatletter
\@ifclassloaded{subfiles}{
  \ifx\mainfile\undefined\else
    \externaldocument{aux/\subfix{\@nameuse{mainfile}}}%
  \fi
}{}
\makeatother

\usepackage{amsmath}
\usepackage{amsfonts}
\usepackage{bbm}
\usepackage{accents}

\usepackage[toc]{appendix}
\usepackage{chngcntr}

\usepackage{graphicx} 	
\usepackage{color} 		
\usepackage[super]{nth} 
\usepackage{float} 		
\usepackage{ragged2e} 
\usepackage{changepage} 
\usepackage{titletoc} 
\usepackage{authoraftertitle} 
\usepackage{datetime} 
\newdateformat{monthyeardate}{\monthname[\THEMONTH] \THEYEAR}

\usepackage{booktabs}
\usepackage{multirow}
\usepackage{hhline} 
\usepackage{makecell} 

\usepackage[bottom,hang,flushmargin]{footmisc}
\usepackage[margin=1in]{geometry}

\usepackage{setspace}

\usepackage[all]{nowidow}

\renewcommand{\arraystretch}{1.7}

\usepackage[colorinlistoftodos]{todonotes} 
\usepackage{mdframed} 		

\usepackage{hyperref}
\hypersetup{colorlinks=true, linkcolor=cyan, urlcolor=cyan, citecolor=cyan}

\AtBeginDocument{

}

\newcommand{\autorefapx}[1]{\hyperref[#1]{Appendix~\autoref*{#1}}}
\newcommand{\autorefapxsec}[1]{\hyperref[#1]{Appendix~\ref*{#1}}}

\usepackage{xcolor}
\usepackage{soul}
\definecolor{codecolor}{RGB}{0,102,153}
\definecolor{codebg}{RGB}{240,240,240}
\sethlcolor{codebg}

\usepackage{titlesec}

\titleformat{\subsubsection}[block]
  {\normalsize\itshape}
  {\thesubsubsection}
  {1em}
  {\ul}

\usepackage{caption}
\usepackage{placeins}

\usepackage[
    backend=biber,
    sorting=nyt, 
    uniquelist=false, 
    hyperref=true, 
    maxcitenames=2, 
    style=apa,
    doi=false, 
    url=false, 
]{biblatex}

\DeclareFieldFormat[article]{title}{\mkbibquote{#1}} 

\DeclareCiteCommand{\parencite}[\mkbibparens]
  {\usebibmacro{prenote}}
  {\printtext[bibhyperref]{%
    \printnames{labelname}%
    \setunit{\printdelim{nameyeardelim}}%
    \printfield{year}%
    \printfield{extradate}}}
  {\multicitedelim}
  {\usebibmacro{postnote}}

\DeclareCiteCommand{\citeauthor}
  {\usebibmacro{prenote}}
  {\printtext[bibhyperref]{%
    \printnames{labelname}}}
  {\multicitedelim}
  {\usebibmacro{postnote}}

\DeclareCiteCommand{\colorcite}
  {\usebibmacro{prenote}}
  {\printtext[bibhyperref]{%
    \printnames{labelname}%
    \setunit{\printdelim{nameyeardelim}}%
    \printfield{year}}%
  }
  {\multicitedelim}
  {\usebibmacro{postnote}}

\usepackage{gelasio}
\usepackage{csquotes}
\newgeometry{left=0.8in, right=0.8in, bottom=0.8in, top=0.8in}
\savegeometry{default_margins}
\title{\textbf{Sorting of Working Parents into Family-Friendly Firms}}

\author{\, \\
    \begin{minipage}[t]{0.24\textwidth}
        \centering
        \textbf{Ross Chu} \vspace{0.5em} \\ \itshape{U.C. Berkeley}
    \end{minipage}%
    \begin{minipage}[t]{0.24\textwidth}
        \centering
        \textbf{Sohee Jeon} \vspace{0.5em} \\ \itshape{Seoul National\\University}
    \end{minipage}%
    \begin{minipage}[t]{0.24\textwidth}
        \centering
        \textbf{Hyun Seung Lee} \vspace{0.5em} \\ \itshape{Seoul National\\University}
    \end{minipage}%
    \begin{minipage}[t]{0.24\textwidth}
        \centering
        \textbf{Tammy Lee} \vspace{0.5em} \\ \itshape{U. Michigan\\(Ann Arbor)}
    \end{minipage}
    \\ [5em]
}

\date{\monthyeardate\today\thanks{
    Ross Chu acknowledges financial support through the National Science Foundation Graduate Research Fellowship (under Grant DGE 2146752). Tammy Lee acknowledges financial support from the National Institute on Aging (under Grant T32AG000221) and the National Institute of Child Health and Human Development (under Grant T32HD007339) through the Population Studies Center (U. Michigan). 
    This work does not represent the views of NSF, NIH, NIA, NICHHD, UC Berkeley, U. Michigan (Ann Arbor), Seoul National University, or Statistics Korea. All statistics reported in this paper represent the analysis sample alone and are not intended to be nationally representative of South Korea. This paper synthesizes three research projects, and we thank faculty and students who gave helpful feedback on individual components: Jake Anderson, Charlie Brown, Luisa Cefala, Konhee Chang, Stefano DellaVigna, Hillary Hoynes, Justin Katz, Kristy Kim, Woojin Kim, Patrick Kline, Alex Mas, Enrico Moretti, Jessica Pan, Ricardo Perez-Truglia, Jesse Rothstein, Dmitry Taubinsky, Christopher Walters, Basit Zafar, and seminar participants at the Berkeley Psychology \& Economics Workshop, Public Finance Seminar at U. Michigan, and ASSA session on the K-Economy. We are grateful to Jungmin Lee for his contributions to early stages of each project, and we thank Seung Hwan Noh for superb research assistance when analyzing the paternity leave mandate. \textcite{tammy_jmp_2025} is a companion paper that analyzes the same data and policy to examine its impacts on fertility decisions.
}
}

\begin{document}

\newgeometry{left=0.8in, right=0.8in, bottom=0.4in, top=0.2in}
\begingroup
\onehalfspacing

\maketitle

\begin{abstract}
    \noindent Using detailed data on workplace benefits linked with administrative registers in Korea, we analyze patterns of separations and job transitions to study how parents sort into family-friendly firms after childbirth.
    We examine two quasi-experimental case studies: 1) staggered compliance with providing onsite childcare, and 2) mandated enrollment into paternity leave at a large conglomerate. 
    In both cases, introducing family-friendly changes attracted more entry by parents who would gain from these benefits, and parents with young children stayed despite slower salary growth.
    We use richer data on a wider range of benefits to show that sorting on family-friendliness mainly occurs through labor force survival rather than job transitions. 
    Most mothers do not actively switch into new jobs after childbirth, and they are more likely to withdraw from the labor force when their employers lack family-friendly benefits.
    We explain these findings with a simple model of sorting that features heterogeneity in outside options and opportunity costs for staying employed, which change after childbirth and vary by gender and family-friendliness at current jobs.
    Taken together, our findings indicate that mothers are concentrated at family-friendly firms not because they switch into new jobs after childbirth, but because they exit the labor force when their employers lack such benefits.
\end{abstract}

\thispagestyle{empty}
\endgroup
\clearpage
\restoregeometry

\begingroup
\singlespacing

\renewcommand{\contentsname}{Table of Contents}
{\small \onehalfspacing \tableofcontents}

\thispagestyle{empty}
\endgroup
\clearpage

\setcounter{page}{1}


\section{Introduction} \label{sec:introduction}

\addcontentsline{toc}{subsection}{Motivation}

    Parenthood is broadly recognized as the primary source of gender pay gaps that remain in the labor market (\colorcite{cortes_children_2023}).
    Gender pay gaps are known to widen significantly after childbirth (\citeauthor{kleven_children_2019} \citeyear{kleven_children_2019,kleven_child_2019}, \citeyear{kleven_child_2025}; \colorcite{kuziemko_mommy_2018}), but what is less understood is where these ``child penalties'' come from.
    A common view is that mothers accept lower wages in exchange for family-friendly benefits, which is true to a certain extent: 
    in the administrative data that we analyze, family-friendly benefits account for 35\% of pay gaps among parents in Korea.\footnote{
        The raw pay gap is 31 log points, which shrinks to 20 log points after controlling for binary indicators of family-friendly benefits with double/debiased machine learning \parencite{chernozhukov_doubledebiased_2018}.
    }
    But how do parents sort into family-friendly firms? Do we expect mothers to interview for new jobs when they are heavily time constrained after childbirth?
    In this paper, we present empirical evidence on how working parents sort into family-friendly firms,
    and we challenge the view that mothers actively pursue job transitions to secure these benefits shortly after childbirth.

    We combine administrative records in Korea with detailed data on workplace benefits to address three research questions.
    First, does providing family-friendly benefits attract more entry by working parents?
    Second, are working parents more likely to stay after introducing such changes?
    Third, are mothers more likely to be at family-friendly firms through job switching or labor force survival?
    These questions have implications for policies that encourage mothers to remain in the labor force after childbirth, which is particularly important when low birth rates put fiscal pressure on welfare programs like social security or public health insurance. 
    
\addcontentsline{toc}{subsection}{Section Summaries}

    The administrative data that we analyze links several statistical registers from Statistics Korea at the individual level.
    The Employment Register contains data on employment and earnings, and data on household composition come from the Population Register.
    The Child Register links all parents to their children, and month-year dates for childbirth and parental leave are obtained through the Parental Leave Register and the Population Dynamics Register.
    The Business Register contains data on employer characteristics, and the government maintains a publicly available list of childcare facilities that allow us to identify employers that provide onsite childcare (\autoref{subsec:childcare_data}).

\begin{figure}[ht]
    \makebox[\textwidth][c]{\includegraphics[width=0.80\textwidth]{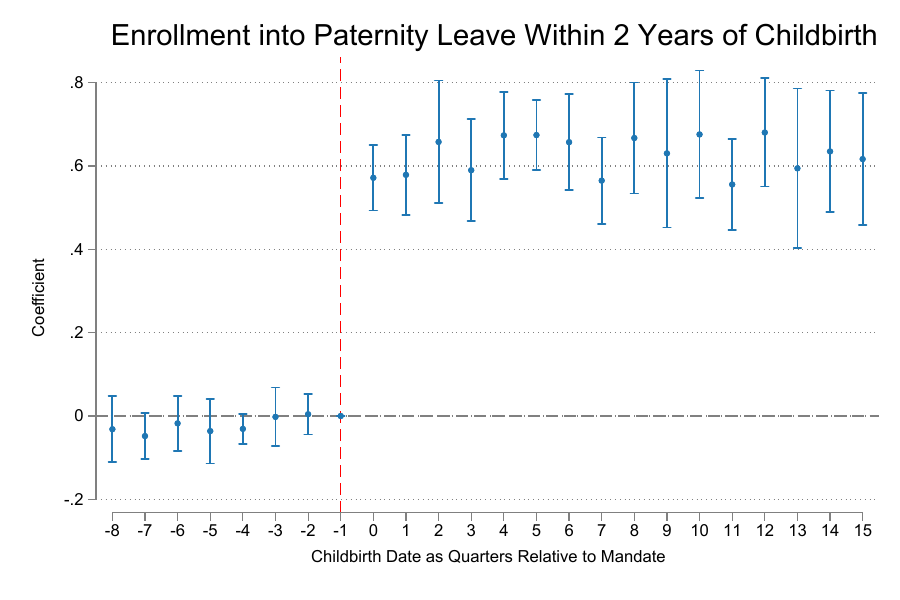}}

    \caption{Impact of Mandated Paternity Leave on Enrollment Among Fathers} \label{fig:patleave_enroll}
    \caption*{\normalsize 
        This figure plots regression coefficients for the impact of mandated paternity leave on enrollment among fathers (discussed in \autoref{subsec:patleave_enroll}). 
        The differences-in-differences (DiD) regression compares fathers at treated and control firms. Treated firms are subsidiaries belonging to the conglomerate that mandated enrollment into paternity leave, and control firms are subsidiaries at other conglomerates that did not implement such policies.
        Observations are fathers $i$ with childbirths at date $t$ (quarterly), and the outcome variable is enrollment in paternity leave within two years of childbirth.
        The regression includes firm fixed effects, calendar time indicators, and interactions between calendar time indicators and an indicator for the treated group.
        The childbirth date (on the horizontal axis) is expressed in quarters relative to enforcement of the mandate. The indicator for $t=-1$ (one quarter prior to the mandate) is omitted to avoid collinearity.
        Each scatter point is the coefficient on the interaction term, and bars denote confidence intervals. 
        Each coefficient is the causal effect at date $t$, interpreted as the difference between fathers at treated and control firms after controlling for calendar time trends and baseline differences across firms.

        \, 

    }
\end{figure}

\begin{figure}[ht]

    \makebox[\textwidth][c]{
        \includegraphics[width=0.48\paperwidth]{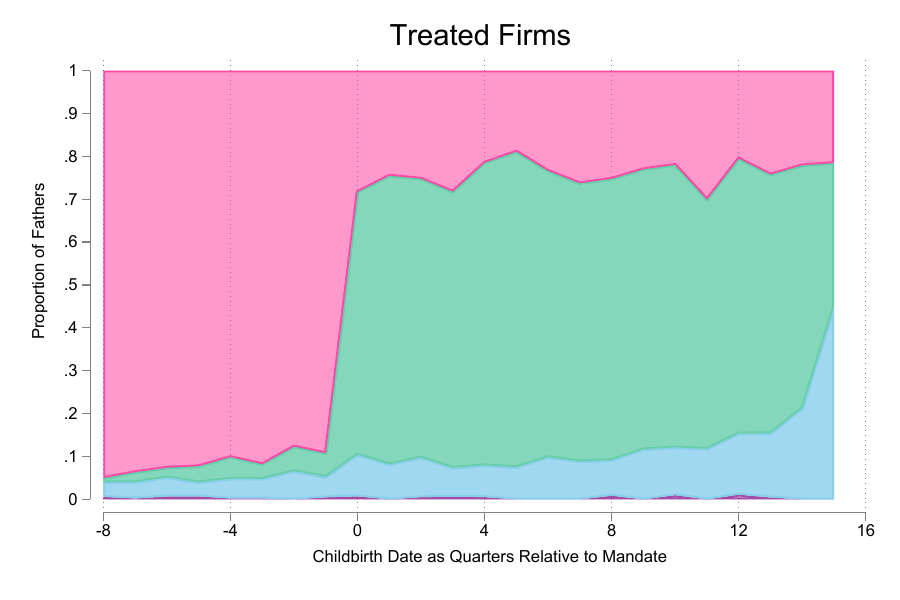}%
        \includegraphics[width=0.48\paperwidth]{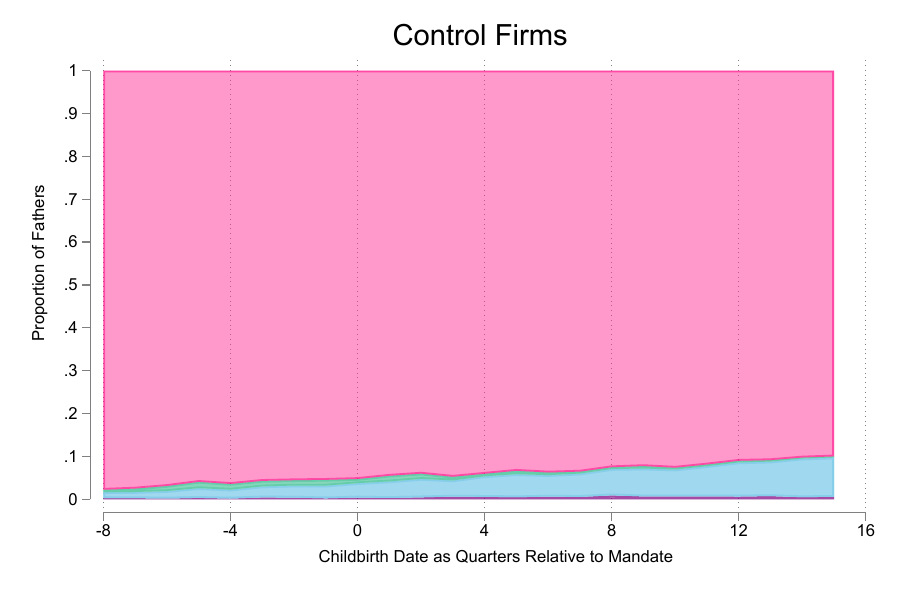}
    }

    \makebox[\textwidth][c]{\includegraphics[width=0.20\paperwidth]{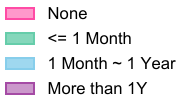}}

    \caption{Distribution of Paternity Leave Durations} \label{fig:patleave_duration}
    \caption*{\normalsize 

        These stacked area plots show the distribution of paternity leave durations for fathers (discussed in \autoref{subsec:patleave_enroll}). 
        The horizontal axis denotes childbirth cohorts, expressed in quarters relative to the enforcement of the mandate. 
        Colors denote the length of paternity leave among fathers in each childbirth cohort. 
        Green denotes 1 month or less, blue denotes between 1-12 months, purple denotes more than 12 months, and pink denotes not enrolling in paternity leave. 
        The height of each area corresponds to the proportion of fathers in the corresponding category. 
        The analysis sample consists of fathers in 1983-95 birth cohorts with childbirths during the sample period, and non-enrollment (in pink) includes fathers who may have been exempt from the mandate due to waivers, part-time hires, or contract employees.

        \, 

    }
\end{figure}

    We use this data to analyze two quasi-experimental case studies: 1) staggered compliance with onsite childcare provision, and 2) mandated enrollment into paternity leave at a large conglomerate.
    The Korean government requires employers to provide onsite childcare since 2006, but compliance has been low due to a lack of enforcement.
    As the government introduced annual penalties to enforce this requirement, compliance increased from 47\% of eligible employers in 2015 to 68\% in 2018 (\autoref{subsec:childcare_background}).
    Our empirical strategy is an event study design with staggered adoption, which we use for two sets of comparisons: one for newly compliant firms and early compliers (pre-2013 vs. 2016-2018), and another for parents and non-parents within treated firms.
    The second quasi-experiment is mandated paternity leave at a large conglomerate, which significantly lowered barriers to enrollment by removing peer pressure.
    All fathers were required to enroll in at least one month of paternity leave within two years of childbirth, which was actively enforced in all subsidiaries under this conglomerate (\autoref{subsec:patleave_background}).
    This mandate was highly effective at increasing enrollment (\autoref{fig:patleave_enroll}), largely driven by fathers who complied at the minimum duration (\autoref{fig:patleave_duration}).
    We evaluate this mandate with a differences-in-differences (DiD) design that compares treated subsidiaries at this conglomerate with subsidiaries at other conglomerates.

\begin{figure}[ht]
    \makebox[\textwidth][c]{\includegraphics[width=0.80\textwidth]{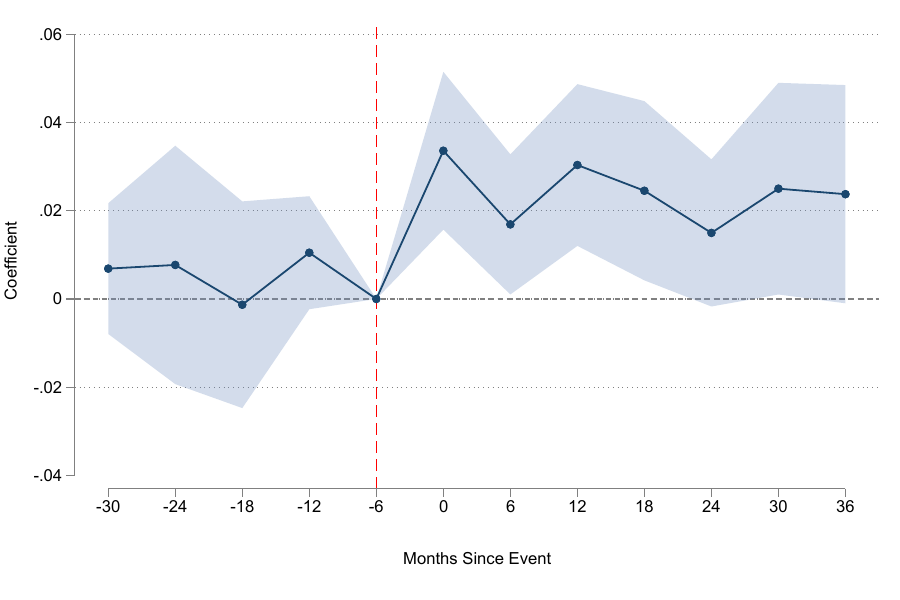}}

    \caption{Impact of Onsite Childcare on the Share of New Hires with Young Children} \label{fig:childcare_entry}
    \caption*{\normalsize 
        This figure plots event study coefficients for the impact of onsite childcare on the share of new hires with young children (discussed in \autoref{subsec:childcare_results}). 
        The regression compares new hires at treated and control establishments. Treated establishments are ``late adopters'' installing onsite childcare facilities between 2016 and 2018, while control establishments are ``early adopters'' installing such facilities before 2013. 
        Observations are firms $j$ at date $t$ (biannual), 
        and observations for each firm are weighted by the number of new hires in 2015. 
        The outcome variable is the share of new hires with young children (age 0-6).
        The regression includes fixed effects for establishments, fixed effects for calendar time, and event time indicators. 
        Event time (on the horizontal axis) is months relative to the installation of onsite childcare for treated establishments, which always equal zero for control establishments. The event time indicator for $t=-6$ is omitted to avoid collinearity.
        Each scatter point is the coefficient on an event time indicator, and shaded areas denote confidence intervals. Each coefficient is the causal effect at event time $t$, interpreted as the difference in outcomes between treated and control establishments after controlling for calendar time trends and baseline differences across establishments. 

        \, 

    }
\end{figure}

\begin{figure}[ht]
    \makebox[\textwidth][c]{\includegraphics[width=0.80\textwidth]{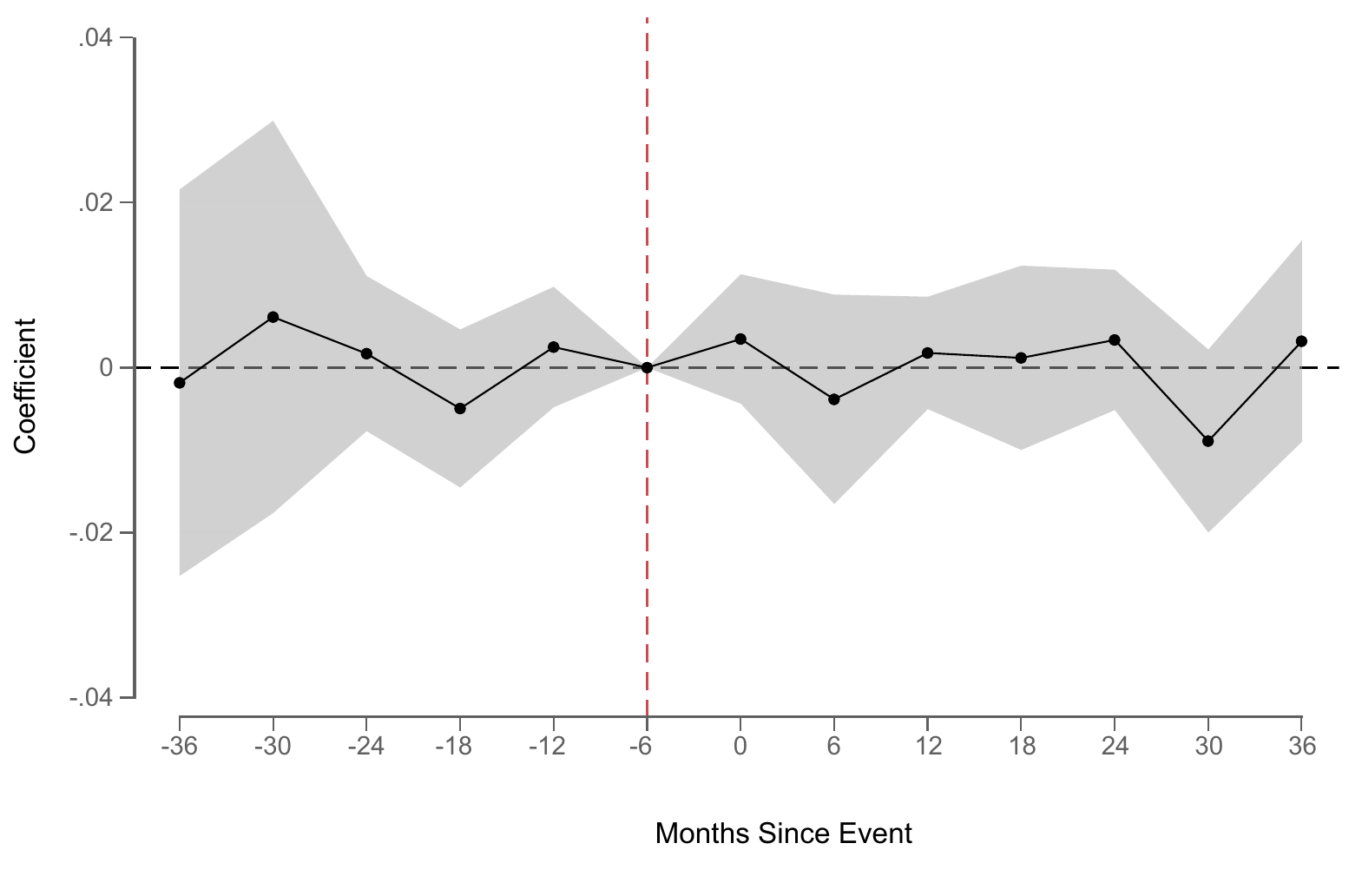}}

    \caption{Impact of Onsite Childcare on Separations for Parents VS Non-Parents} \label{fig:childcare_separation}
    \caption*{\normalsize 
        This figure plots event study coefficients for the impact of onsite childcare on separation rates (discussed in \autoref{subsec:childcare_results}). 
        The regression compares parents and non-parents within treated establishments installing onsite childcare facilities between 2016 and 2018.
        Each coefficient corresponds to the difference between parents and non-parents at each point in event time. 
        Observations are employees $i$ at date $t$.
        The outcome variable is employee separation at time $t$, which is defined for those employed until time $t-1$. The separation indicator equals 1 if separating at time $t$, and equals 0 otherwise. 
        The regression includes time-varying attributes at the person level (age, tenure, tenure squared), event-time indicators, a parent indicator, and interactions between event-time and parent indicators. 
        Event time (on the horizontal axis) is months relative to the installation of onsite childcare, and the indicator for $t=-6$ is omitted to avoid collinearity.
        Scatter points are coefficients on interaction terms, and shaded areas denote confidence intervals. 
        Each coefficient is the causal effect at event time $t$, interpreted as the difference between parents and non-parents after controlling for age and tenure. 

        \, 



    }
\end{figure}

\begin{figure}[ht]
    \makebox[\textwidth][c]{\includegraphics[width=0.80\textwidth]{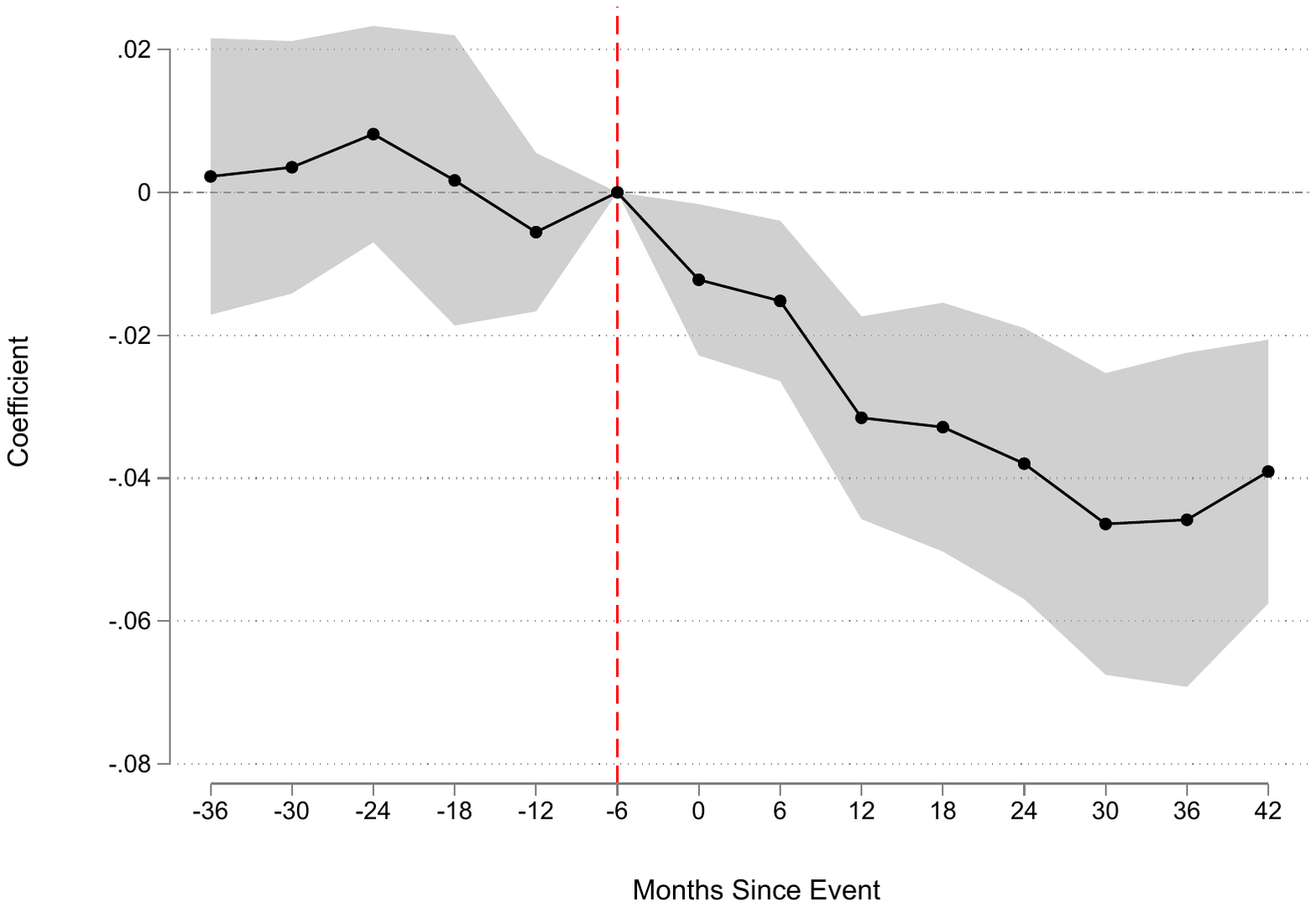}}

    \caption{Impact of Onsite Childcare on Log Salaries for Working Parents} \label{fig:childcare_salary}
    \caption*{\normalsize 
        This figure plots event study coefficients for the impact of onsite childcare on log salaries for working parents (discussed in \autoref{subsec:childcare_results}). 
        The regression compares parents and non-parents within treated establishments installing onsite childcare facilities between 2016 and 2018.
        Observations are employees $i$ at date $t$.
        The outcome variable is log salary, which divides total earnings by the number of months worked during the year.
        The regression includes person fixed effects, time-varying attributes at the person level (age, tenure, tenure squared), linear time trends specific to each income rank, event-time indicators, and interactions between event-time and parent indicators. 
        Event time (on the horizontal axis) is months relative to the installation of onsite childcare, and the indicator for $t=-6$ is omitted to avoid collinearity.
        Scatter points are coefficients on interaction terms, and shaded areas denote confidence intervals. 
        Each coefficient is the causal effect at event time $t$, interpreted as the difference between parents and non-parents after controlling for person-level attributes and income-specific time trends.

        \, 

    }
\end{figure}

\begin{figure}[ht]
    \makebox[\textwidth][c]{\includegraphics[width=0.80\textwidth]{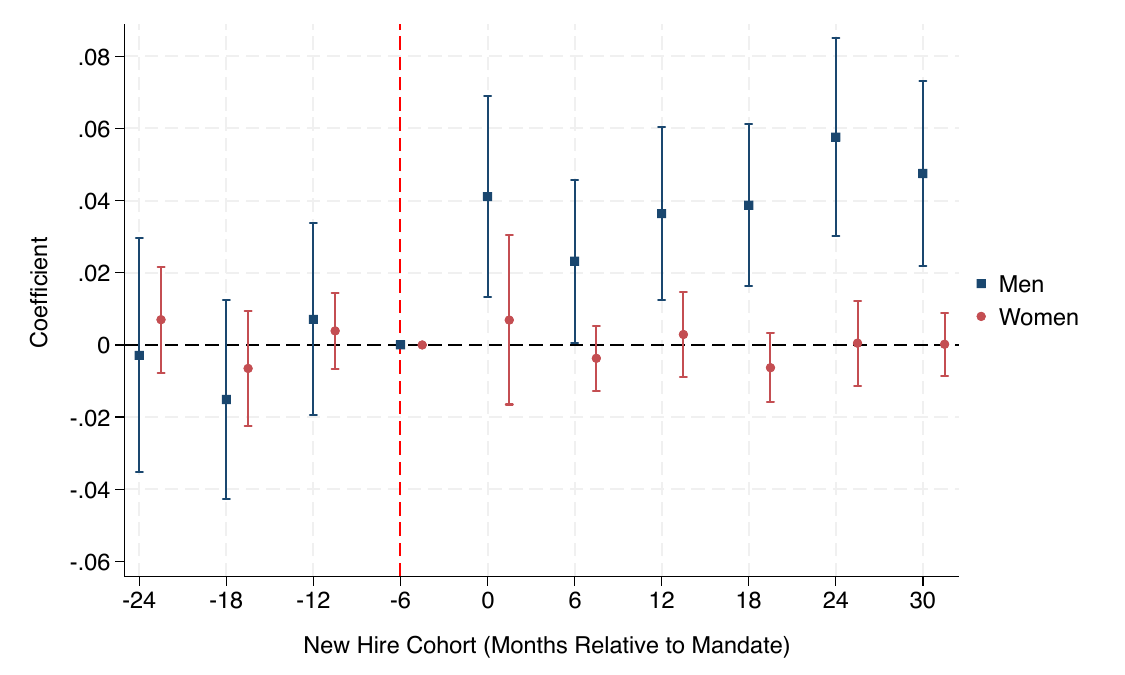}}

    \caption{Impact of Mandated Leave on the Share of New Hires with Impending Childbirths} \label{fig:patleave_entry}
    \caption*{\normalsize 
        This figure plots regression coefficients for the impact of mandated paternity leave on the share of new hires with impending childbirths (discussed in \autoref{subsec:patleave_results}).
        The differences-in-differences (DiD) regression compares new hires at treated and control firms. Treated firms are subsidiaries belonging to the conglomerate that mandated enrollment into paternity leave, and control firms are subsidiaries at other conglomerates that did not implement such policies.
        Observations are new hire cohorts at date $t$ for firm $j$ (biannual).
        The outcome is the share of new hires with impending childbirths, defined as having childbirth after being hired. The denominator is the number of new hires aged 25-40 in each cohort, and the numerator is the number of men (in blue) and women (in red) with childbirths after their hire dates.
        The regression includes firm fixed effects, calendar time fixed effects, and interactions between calendar time indicators and an indicator for the treated group. 
        The hire date (on the horizontal axis) is expressed in months relative to the enforcement of the mandate. The indicator for $t=-6$ (six months prior to the mandate) is omitted to avoid collinearity.
        Each scatter point is the coefficient on the interaction term, and bars denote confidence intervals. 
        Each coefficient is the causal effect at date $t$, interpreted as the difference between treated and control firms after controlling for calendar time trends and baseline differences across firms.

        \, 

    }
\end{figure}

\begin{figure}[ht]
    \makebox[\textwidth][c]{\includegraphics[width=0.80\textwidth]{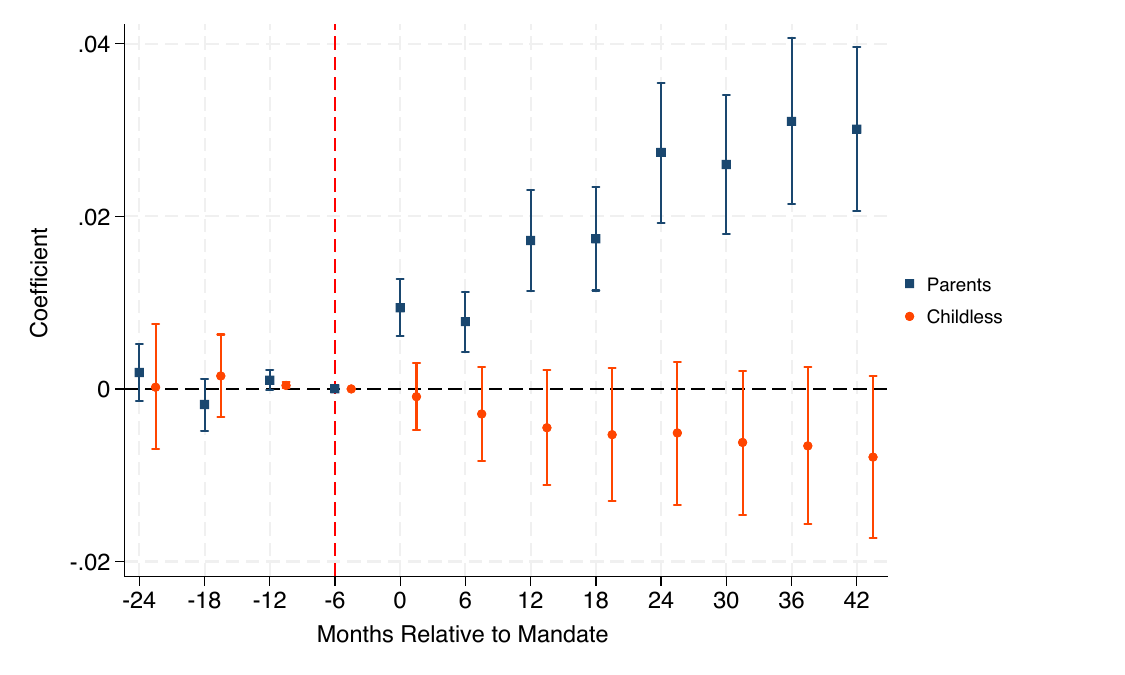}}

    \caption{Impact of Mandated Paternity Leave on Employee Retention} \label{fig:patleave_retention}
    \caption*{\normalsize 
        This figure plots regression coefficients for the impact of mandated paternity leave on employee retention (discussed in \autoref{subsec:patleave_results}).
        The differences-in-differences (DiD) regression compares employees at treated and control firms. Treated firms are subsidiaries belonging to the conglomerate that mandated enrollment into paternity leave, and control firms are subsidiaries at other conglomerates that did not implement such policies.
        Observations are firms $j$ at date $t$ (biannual). The outcome variable is the male share of staying parents (in blue) and non-parents (in orange). 
        For these shares, the denominator is the number of men (aged 25-40) hired before the mandate, and the numerator is the number of fathers and childless men remaining at each date. Parents (in blue) are fathers with young children (age 0-2), and non-parents (in orange) remain childless until the end of the sample period. 
        The regression includes firm fixed effects, calendar time fixed effects, and interactions between calendar time indicators and an indicator for the treated group. 
        Calendar dates (on the horizontal axis) are expressed in months relative to the introduction of the mandate. The indicator for $t=-6$ (six months prior to the mandate) is omitted to avoid collinearity.
        Each scatter point is the coefficient on the interaction term, and bars denote confidence intervals. 
        Each coefficient is the causal effect at date $t$, interpreted as the difference between treated and control firms after controlling for calendar time trends and baseline differences across firms.

        \, 

    }
\end{figure}

    Mandating paternity leave attracted more entry by future fathers who would benefit from this change. The mandate resulted in a rising share of newly hired men with childbirths after their hire dates (\autoref{fig:patleave_entry}), and men were more likely to stay and have children after the mandate (\autoref{fig:patleave_retention}).
    Similarly, introducing onsite childcare attracted more entry by parents with young children, who benefit the most from these services (\autoref{fig:childcare_entry}).
    Parents were not more likely to separate after onsite childcare (\autoref{fig:childcare_separation}), despite slower salary growth compared with non-parents (\autoref{fig:childcare_salary}).

    We use richer data on a wider range of benefits to examine whether sorting in the broader labor market is primarily driven by job switching or labor force survival.
    We augment the administrative data with two crowdsourced datasets on workplace benefits.
    The first dataset comes from a platform with crowdsourced details on benefits offered at each company, which is submitted via surveys and verified by human resource representatives.
    The second dataset comes from a platform with crowdsourced ratings on each company. These 5-point Likert scale ratings are similar to Glassdoor.com, but they additionally benefit from corporate email verification and predominant usage in the Korean job market.
    Using this data, we mark employers as ``family-friendly'' if they: i) have work-life balance ratings above 4, ii) offer benefits regarding fertility, pregnancy, or childcare, or iii) provide family-friendly workplace arrangements (\autoref{subsec:separation_data}).

\begin{figure}[ht]

    \makebox[\textwidth][c]{
        \includegraphics[width=0.45\paperwidth]{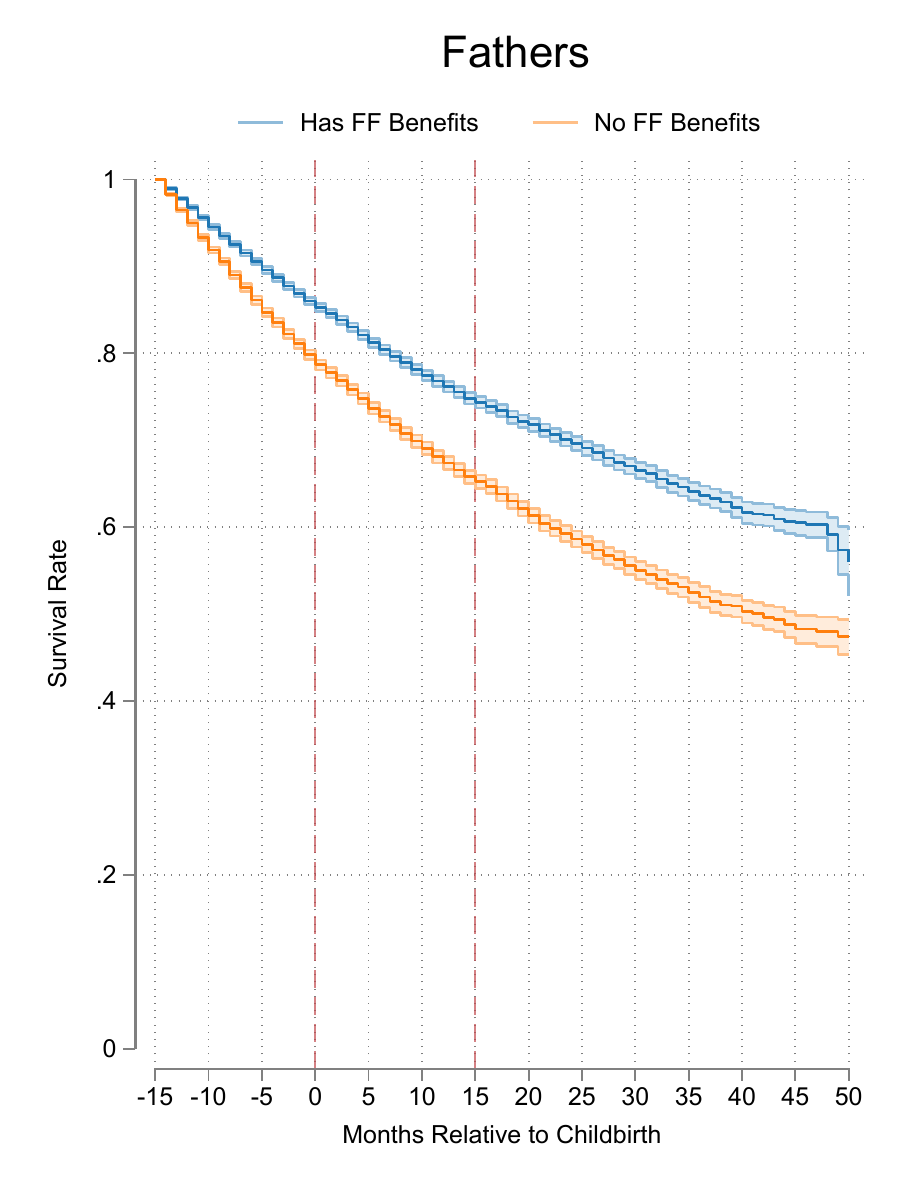}
        \includegraphics[width=0.45\paperwidth]{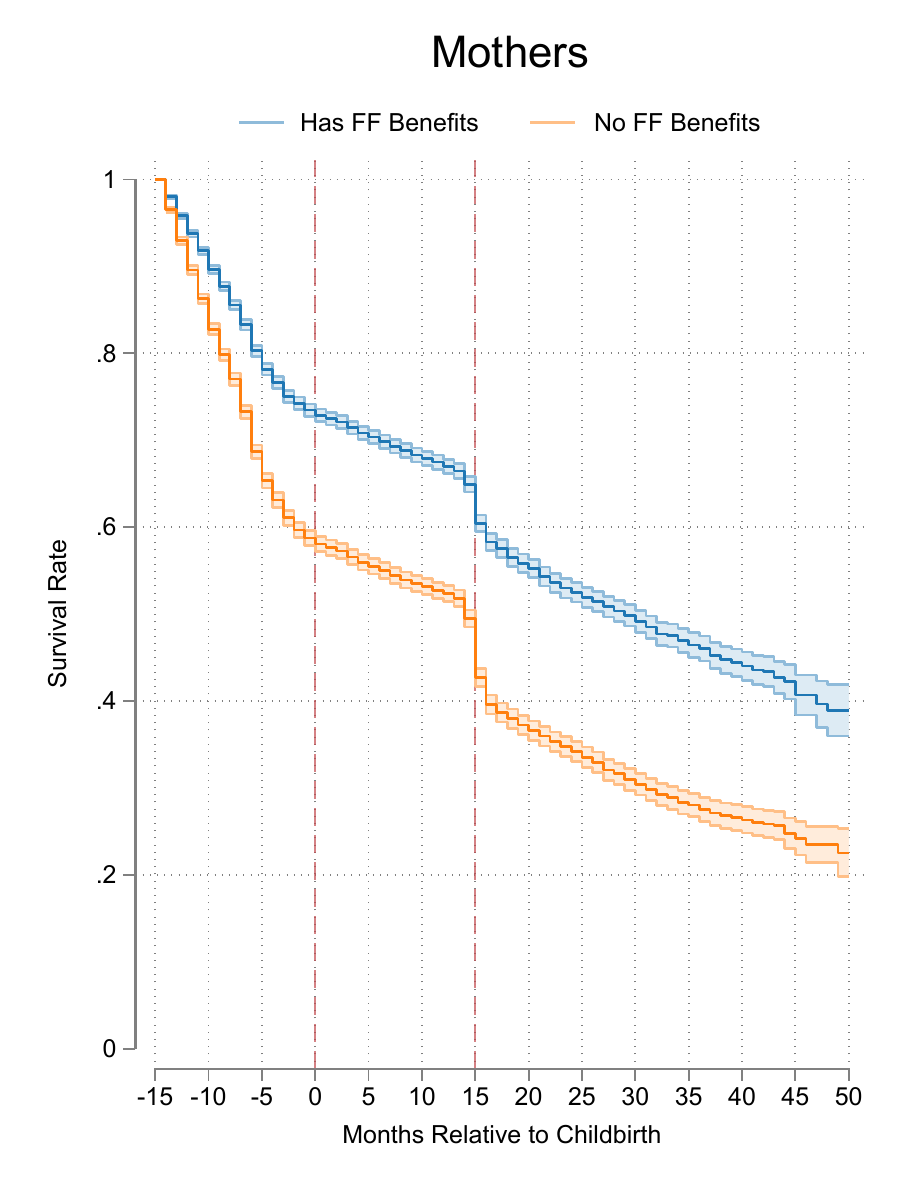}
    }

    \caption{Survival Rates at Firms With VS Without Family-Friendly Benefits} \label{fig:survival_friendly}
    \caption*{\normalsize 

        The Kaplan-Meier survival function shows the probability of remaining with the firm at each point in time (discussed \autoref{subsec:separation_survival}). The horizontal axis denotes months relative to childbirth, and the vertical axis denotes the probability of remaining with the firm. Observations are job spells for parents employed with the firm 15 months prior to childbirth ($t=-15$), which record tenure accumulated since $t=-15$ at separation. Job spells belong to parents with childbirth during their tenure. The survival function accounts for right-censoring due to unobserved separations. 

        \,

        The left panel corresponds to fathers, and the right panel corresponds to mothers. Blue lines correspond to firms with family-friendly benefits (``Has FF Benefits''), and orange lines correspond to firms without such benefits (``No FF Benefits''). The first vertical line at $t=0$ corresponds to childbirth, and the second vertical line at $t=15$ corresponds to the end of paid parental leave (3 months of pregnancy leave + 12 months of maternity leave). Shaded areas denote confidence intervals.
    }
\end{figure}

\begin{figure}[ht]

    \makebox[\textwidth][c]{\includegraphics[width=0.47\paperwidth]{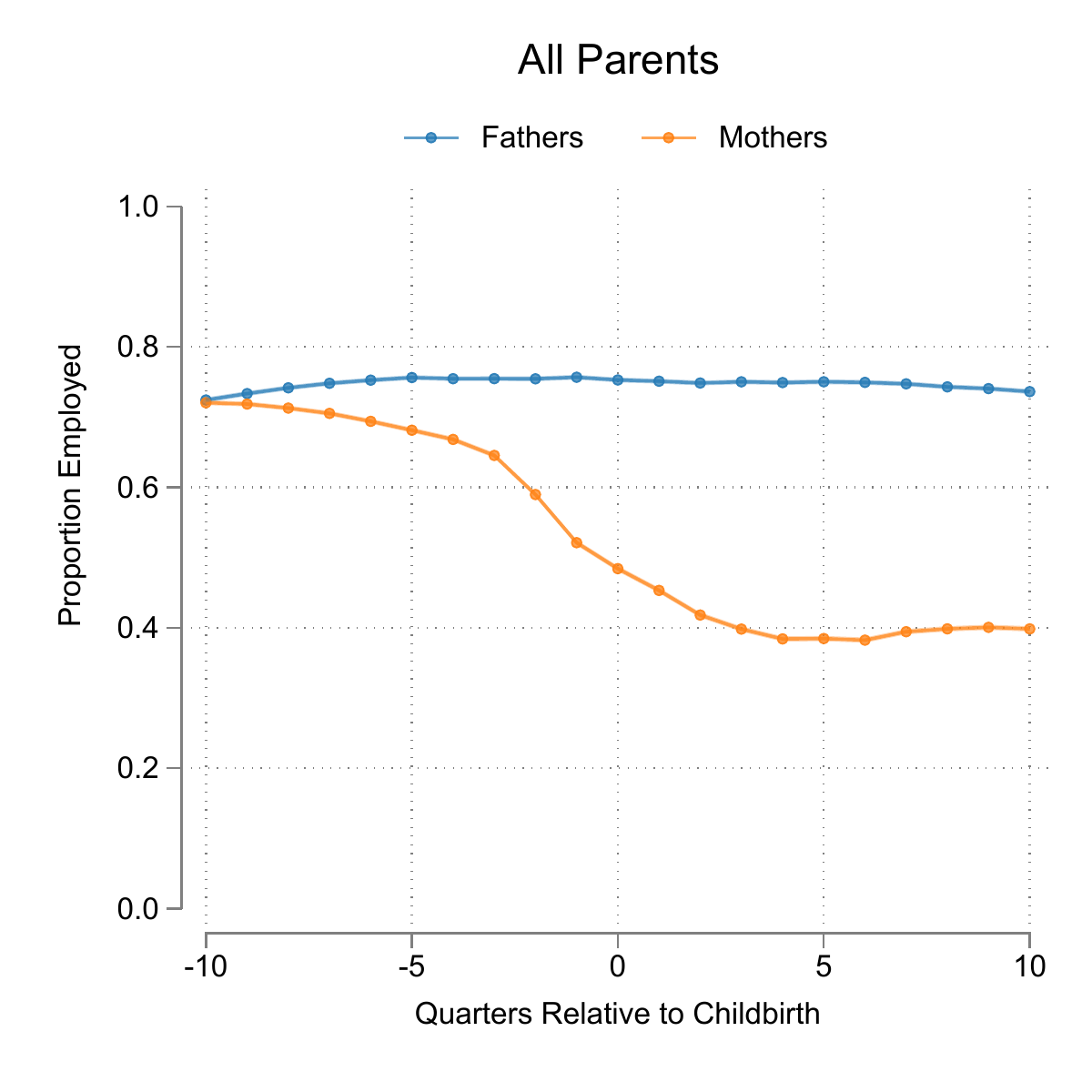}}

    \makebox[\textwidth][c]{%
        \includegraphics[width=0.47\paperwidth]{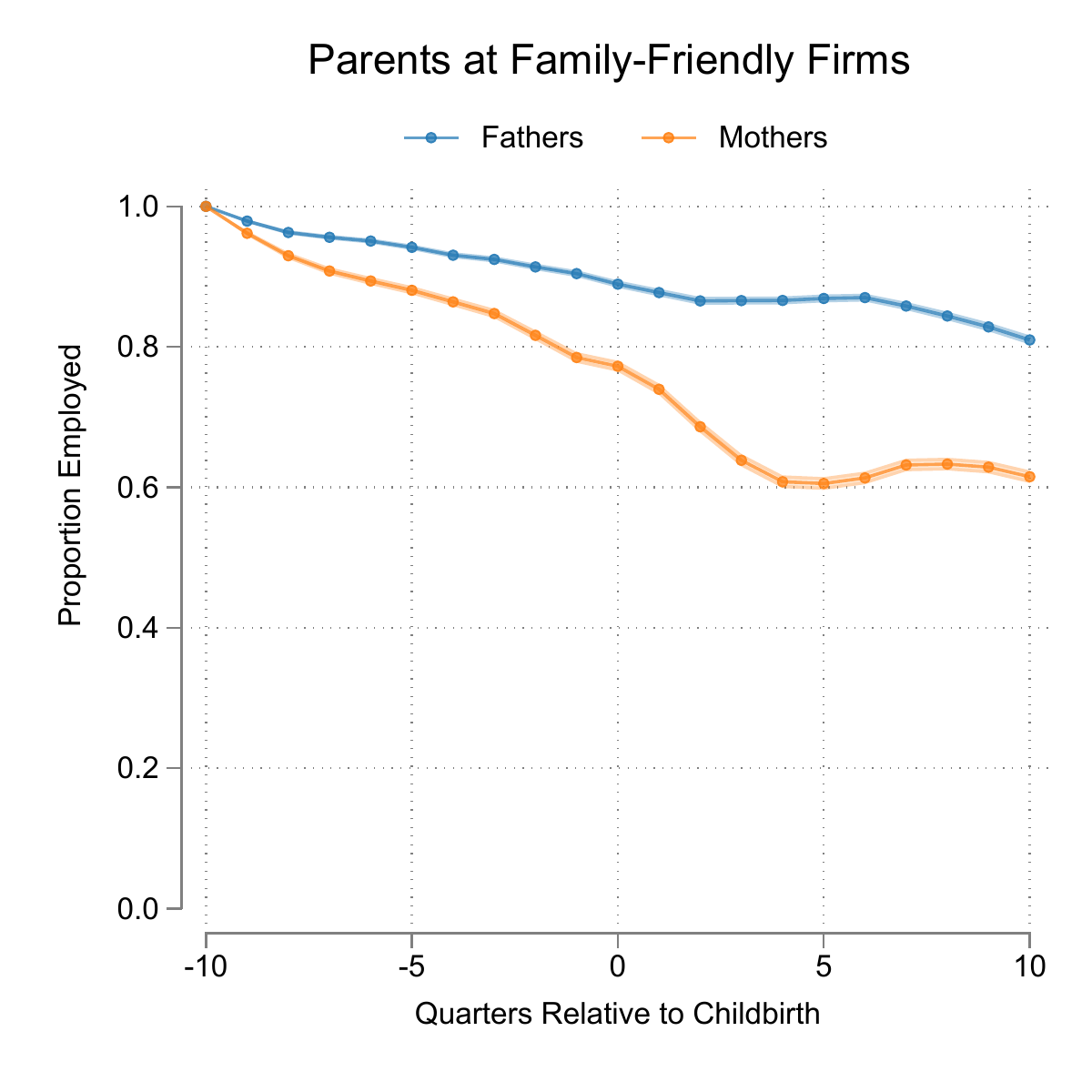}%
        \includegraphics[width=0.47\paperwidth]{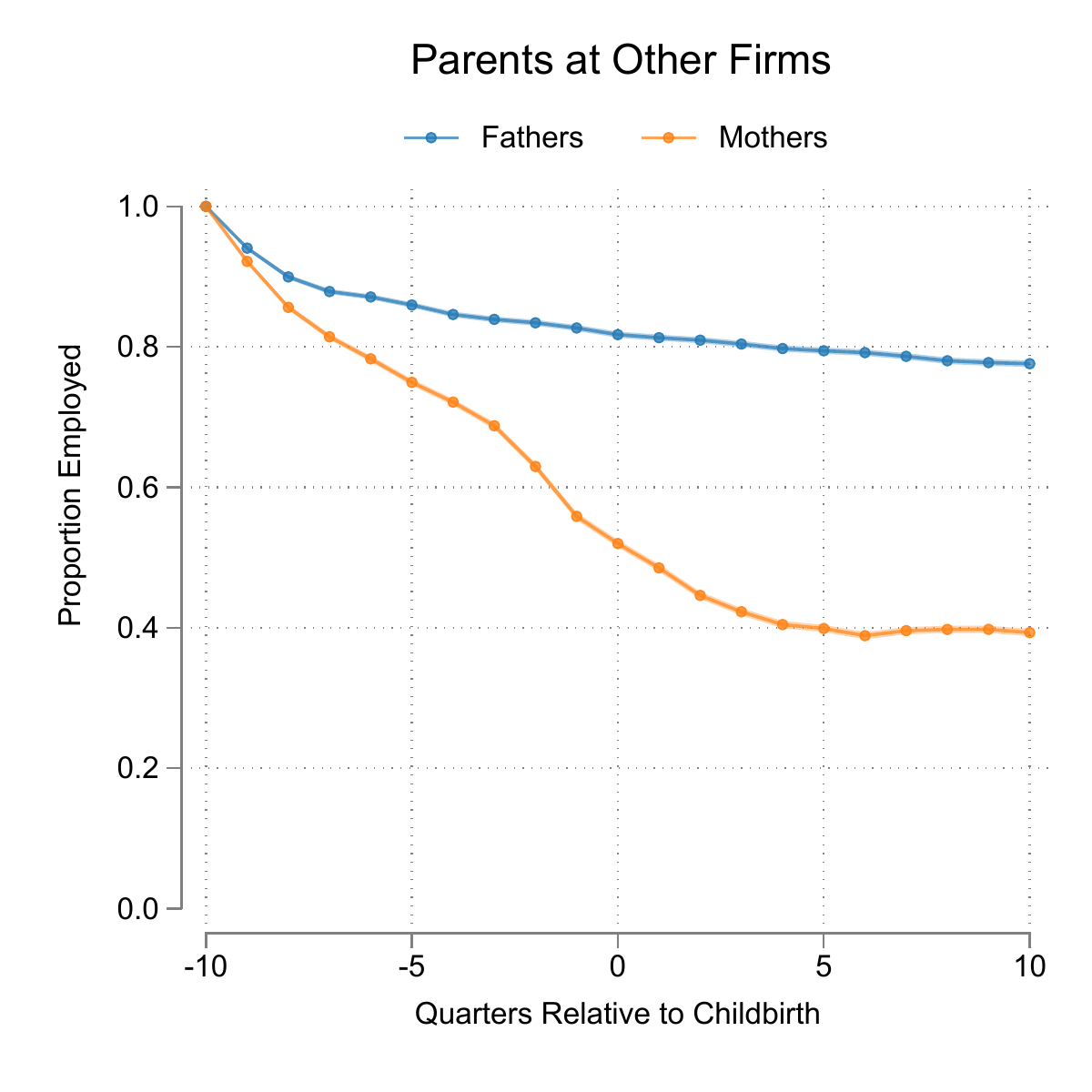}
    }
    \caption{Share of Parents Remaining in the Labor Force} \label{fig:share_employed}
    \caption*{\normalsize 
        The top panel plots employment rates for a balanced panel of all parents, and bottom panels plot employment rates for parents at family-friendly firms (bottom left) and other firms (bottom rightl) 10 quarters prior to childbirth (discussed in \autoref{subsec:separation_share}). The sample consists of parents with childbirths between July 2017 and June 2018, which corresponds to a balanced panel for 10 quarters before and after childbirth. Employment includes part-time and full-time jobs with unemployment insurance coverage, and family-friendliness is zero-coded for firms with missing data on workplace benefits. Shaded areas denote confidence intervals.
    }
\end{figure}

\begin{figure}[ht]

    \makebox[\textwidth][c]{
        \includegraphics[width=0.48\paperwidth]{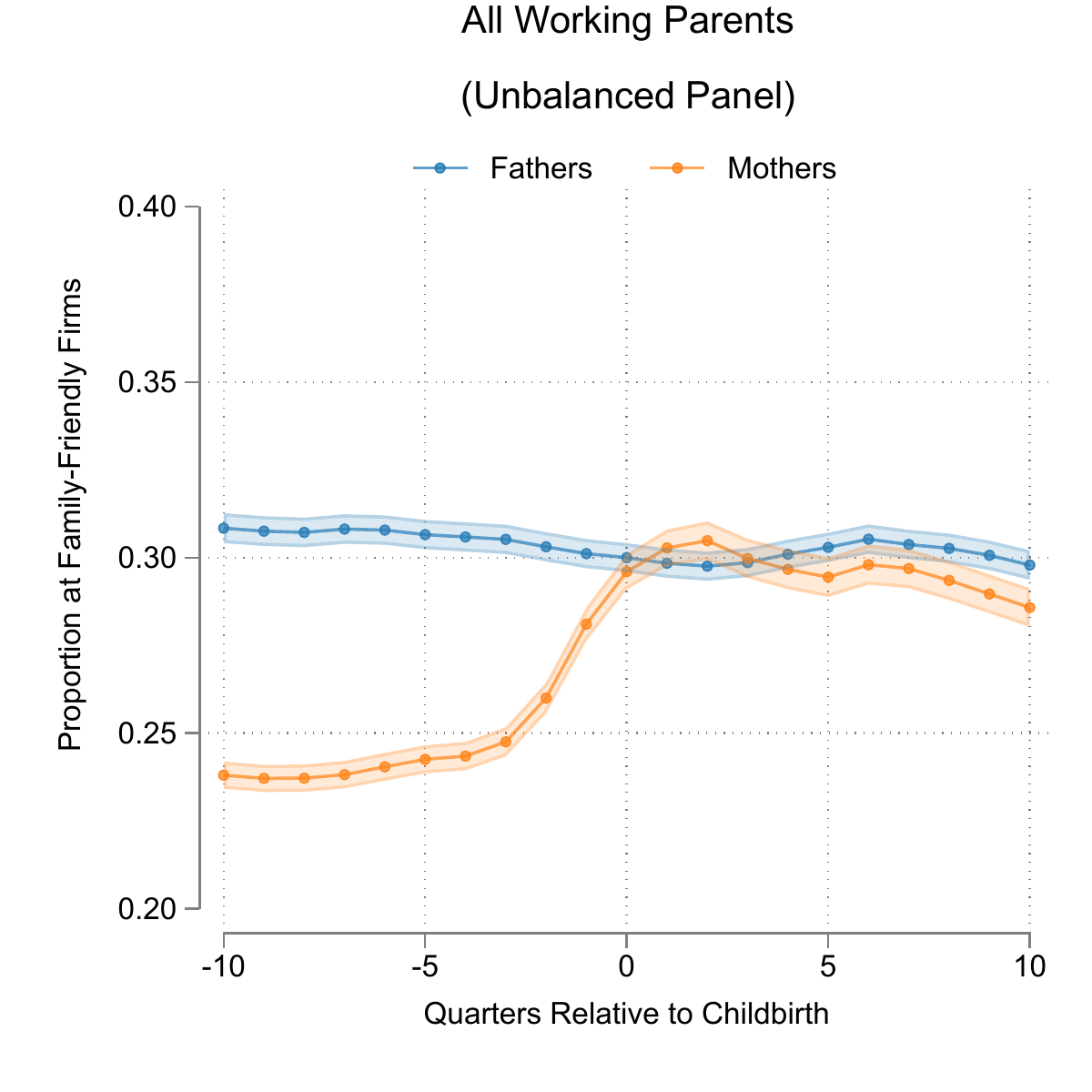}
        \includegraphics[width=0.48\paperwidth]{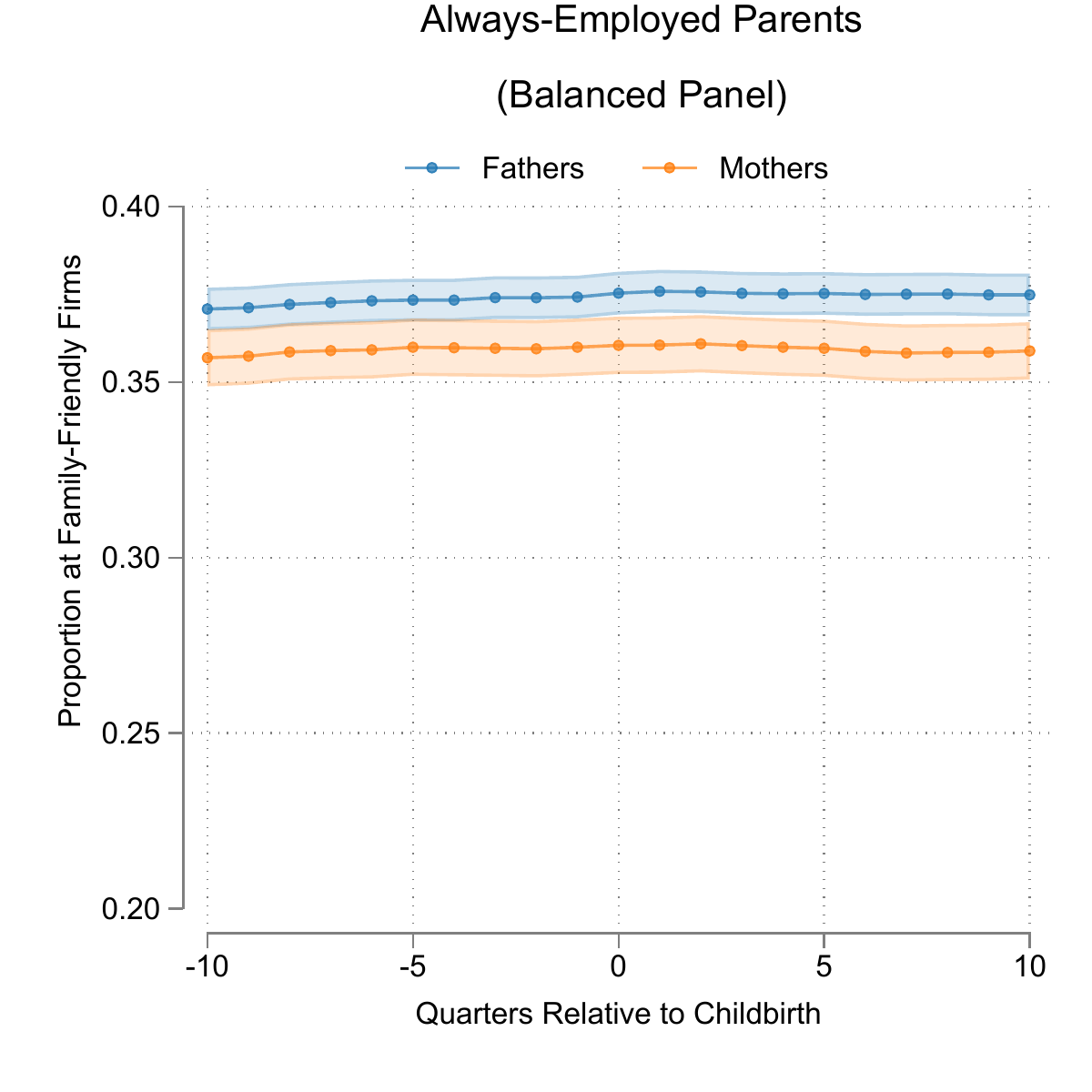}
    }

    \caption{Share of Working Parents at Family-Friendly Firms} \label{fig:share_friendly}
    \caption*{\normalsize
        These figures plot the share of working parents at family-friendly firms before and after childbirth (discussed in \autoref{subsec:separation_share}). The figure on the left is for an unbalanced panel of parents who remain in the labor force, and the figure on the right is for a balanced panel of parents who are always employed. The sample consists of parents with childbirths between July 2017 and June 2018. Employment includes part-time and full-time jobs at any firm with unemployment insurance (UI) coverage, and family-friendliness is zero-coded for firms with missing data on workplace benefits. Shaded areas denote confidence intervals.
    }
\end{figure}

    Using this data, we document empirical evidence that sorting on family-friendliness mainly occurs through labor force survival rather than job transitions.
    Parents are less likely to separate from firms with family-friendly benefits (\autoref{fig:survival_friendly}), and cross-industry differences and amenities unrelated to family-friendliness can only partially explain higher survival rates at family-friendly firms (\autoref{tab:hazard_regressions}).
    Mothers are more likely to be at family-friendly firms after childbirth (\autoref{fig:share_friendly}), but this is not because they actively switch into new jobs (only 2.6\% do so after childbirth).
    Instead, mothers are concentrated at family-friendly firms because they withdraw from the labor force when their employers lack such benefits (\autoref{fig:share_employed}).

    We explain these findings with a simple model where employed parents make one of three decisions: 1) stay at the current job, 2) switch into a new job, or 3) transition into non-employment.
    Choice probabilities on each decision are shaped by unobserved heterogeneity in labor productivity, non-wage amenities, and non-employment values.
    We empirically measure choice probabilities before and after childbirth  (\autoref{fig:choice_prob}), which align with earlier findings on separations, labor force survival, and higher concentrations of mothers at family-friendly firms.
    For each of these findings, we intuitively explain changes in choice probabilities after childbirth with changes in unobserved heterogeneity that vary by gender and family-friendliness at current jobs.

    These findings have implications on how governments should design policies that encourage mothers to remain in the labor force after childbirth.
    For developed countries with low fertility, doing so is important for broadening the tax base to ease financial burden on social security and public health insurance.
    Most mothers are severely time constrained after childbirth, and they are often tied to their current jobs as most lack the capacity to interview for new positions.
    Since mothers are more likely to leave the labor force when their employers lack family-friendly benefits, governments could prevent labor force withdrawals by targeting their resources to this segment of the labor market.
    For example, governments might mandate family-friendly benefits for larger employers while stepping in for smaller employers that cannot effectively offer them at scale (e.g. mandating onsite childcare for larger firms while expanding public childcare for smaller firms).

\addcontentsline{toc}{subsection}{Literature Review}

    We discuss our contributions to two strands of closely related literature.
    First, this paper augments pre-existing evidence for sorting on non-wage amenities.
    Several papers show that workers tend to be in jobs with amenities that they value highly:
    \textcite{mas_valuing_2017} find that workers with work-from-home arrangements value this benefit twice as much as workers without such benefits, and \textcite{corradini_collective_2025} report that firms introducing female-friendly amenities attracted more entry by women (see \colorcite{mas_non-wage_2025} for a thorough review of empirical evidence on sorting).\footnote{
        However, there is also evidence on the contrary.
        For certain amenities, there are unanticipated relationships between workers’ willingness-to-pay (WTP) and their likelihood of having them.
        \textcite{mas_valuing_2017} find that workers with high WTP to avoid irregular work schedules are not less likely to be in those jobs, and \textcite{lachowska_work_2023} find that workers who prefer longer working hours are not more likely to be at firms with longer hours.
    } We complement \textcite{corradini_collective_2025} with two quasi-experimental cases showing that firms introducing \underline{family-friendly} benefits attracted more entry by parents and lowered their separations.
    Furthermore, our observational analyses provide direction on how these findings generalize to the broader labor market: mothers are highly unlikely to switch jobs after childbirth, and the rising share of mothers at family-friendly jobs is primarily driven by labor force withdrawal when employers lack such benefits.

    Second, this paper broadens empirical evidence on career changes after childbirth, which is important to the literature on gender pay gaps and motherhood penalties.
    Prior studies show that mothers are willing to pay more for family-friendly job attributes, and they are more likely to be in family-friendly jobs after childbirth.\footnote{
        Work-From-Home arrangements \parencite{maestas_value_2023}, predictable work schedules \parencite{mas_valuing_2017}, part-time hours \parencite{wasserman_hours_2023, wiswall_preference_2018}, and job stability \parencite{wiswall_preference_2018}.
    } 
    While many studies infer the``family-friendliness'' of jobs indirectly through industries or coworker demographics\footnote{
        Examples of family-friendly jobs include public-sector positions \parencite{pertold-gebicka_employment_2016, kleven_children_2019}, industries with higher female ratios \parencite{cha_overwork_2013}, and higher female/parental share among colleagues \parencite{hotz_parenthood_2018}. \textcite{hotz_parenthood_2018} uses job transitions to propose a model-based index for family-friendliness, which is complementary to these approaches.
    }, our contribution is to directly identify firms that provide family-friendly benefits.
    Like prior studies, we use specific amenities to exploit quasi-experimental variation in narrow contexts (e.g. expansion of onsite childcare).
    However, our study additionally benefits from linking administrative data with detailed workplace amenities, which yields a holistic measure of family-friendliness that spans a wider range of workplace attributes in the overall labor market.

    \,

\noindent The rest of this paper is organized as follows: \vspace{-6pt}

\begin{adjustwidth}{-0.6em}{-0.1em} \vspace{1em}
\begin{itemize}
    \item \autoref{sec:childcare} analyzes entries and separations at firms that introduce onsite childcare.
    \item \autoref{sec:patleave} examines similar evidence at firms with mandated paternity leave for fathers.
    \item \autoref{sec:separation} uses a wider range of benefits to document patterns of job switching and labor force survival.
    \item \autoref{sec:model} interprets empirical results with a simple model of sorting after childbirth.
    \item \autoref{sec:conclusion} concludes by discussing policy implications.
\end{itemize}
\end{adjustwidth}

\clearpage

\section{Staggered Adoption of Onsite Childcare} \label{sec:childcare}

    We examine two quasi-experimental case studies of sorting into family-friendly workplaces: 1) staggered compliance with onsite childcare provision, and 2) mandated enrollment into paternity leave at a large conglomerate.
    We begin by discussing onsite childcare, which shows that working parents are responsive to the availability of family-friendly benefits.

\subsection{Background on Onsite Childcare} \label{subsec:childcare_background}

    As one of many measures to address declining fertility in Korea, the government pursued multiple avenues for expanding access to childcare services.
    Household costs for childcare services are heavily subsidized by the government. Subsidies cover nearly 90\% of enrollment costs for childcare services, and the average household spends around 150k KRW per month on private childcare (1,000 KRW $\approx$ \$1 in purchasing power).
    There are two alternatives to private childcare that are strongly preferred by parents and often have long waiting lists.
    The first is public childcare, which is effectively free, benefits from government accountability, and has a higher bar on teacher credentials.
    The second alternative is onsite childcare provided by employers, which has longer hours of service (to accommodate overtime hours), higher qualifications for teachers, and the convenience of coordinating pick-up and drop-off.
    Despite strong demand for such services, onsite childcare is not readily accessible to parents. While 70\% of children aged 0-6 were enrolled in some form of childcare in 2017, only 4-5\% were enrolled in onsite childcare.

\begin{figure}[ht]
    \makebox[\textwidth][c]{\includegraphics[width=0.80\textwidth]{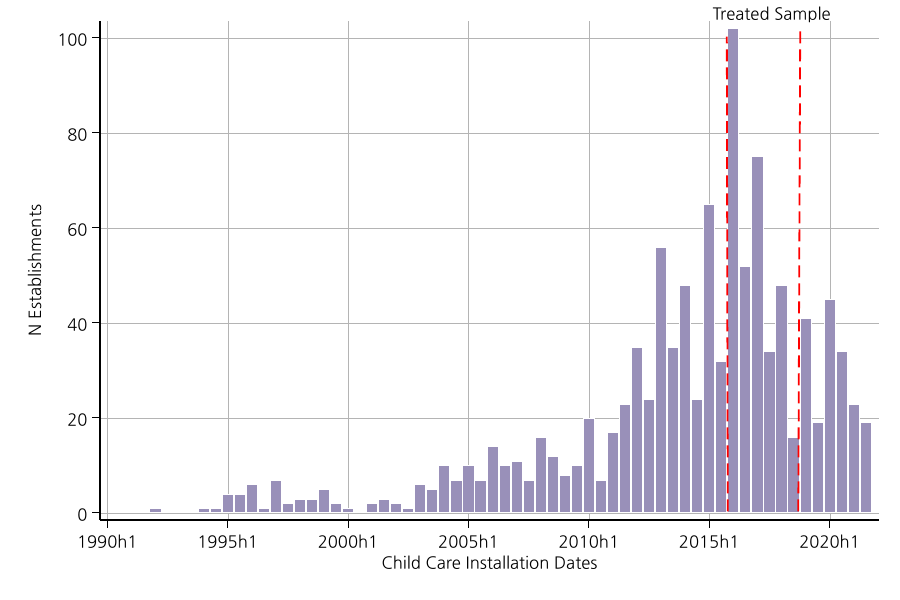}}

    \caption{Distribution of Installation Dates for Onsite Childcare Facilities} \label{fig:childcare_dates}
    \caption*{\normalsize 
        This figure shows the distribution of installation dates for onsite childcare facilities for establishments eligible under the policy (discussed in \autoref{subsec:childcare_background}).

        \,

    }
\end{figure}

    The Korean government requires employers to provide onsite childcare since 2006, but compliance has been low due to a lack of enforcement.
    The government introduced two major changes in 2015 to increase compliance.
    First, employers were no longer allowed to offer cash transfers in lieu of onsite childcare. Second, employers were charged an annual penalty of 150 million KRW for non-compliance with childcare provision ($\approx$ \$150k in purchasing power). As these changes were enforced in 2016, compliance increased from 47\% of employers in 2015 to 68\% of employers in 2018.
    \autoref{fig:childcare_dates} shows the distribution of installation dates for onsite childcare facilities.
    Our empirical analysis compares newly compliant firms with early compliers: newly compliant firms introduced onsite childcare between 2016 and 2018, while early compliant firms introduced such services prior to 2013.

    The policy mandate extends to establishments with more than 500 employees or 300 women, which is applicable to 750-850 establishments each year.
    The government provides financial support for building childcare facilities by subsidizing 30\% of teacher wages and 60\% of installation costs.
    Employer subsidies are capped at 262 million KRW ($\approx$ 58\% of installation costs averaging at 448 million KRW). Since penalties for non-compliance are 150 million KRW annually, employers are financially incentivized to comply with the policy mandate.

\subsection{Linked Administrative Data} \label{subsec:childcare_data}

    The administrative data that we analyze links several statistical registers from Statistics Korea at the individual level.
    Matched employer--employee data come from the Employment Register, which provide data on employment and earnings for workers registered with the unemployment insurance agency between 2015 and 2020 (which notably excludes freelancers, gig workers, temporary contractors, and self-employed individuals).
    The Population Register contains data on household composition (occupants and their relationship to the head of household), and the Child Register links all children with their parents. We use this data to track childbirth status for each adult.

    The Business Register contains data on employer characteristics, and compliance with onsite childcare provision is determined through a publicly available list of childcare facilities maintained by the government.
    The finest level of granularity in matched employer-employee data is a business registration number, which can contain multiple establishments.
    Since eligibility criteria for onsite childcare is based on the size of an establishment, the analysis sample consists of business registration numbers with exactly one establishment in a commuting zone with 500+ employees or 300+ women.\footnote{Commuting zones are defined as administrative provinces (si-do level), but we combine Seoul, Incheon, and Gyeonggi provinces into a single commuting zone considering their geographical proximity.}
    This restriction results in a 33\% coverage for treated establishments (newly compliant firms) and 39\% coverage for control establishments (early compliant firms).
    We link childcare facilities to establishments if both of them are in the same residential district, and we exclude employees at ineligible establishments within compliant businesses (inferred through their home addresses).

\subsection{Event Study with Staggered Compliance} \label{subsec:childcare_strategy}

    The empirical strategy is an event study analysis that generalizes differences-in-differences (DiD) estimation for multiple time periods and event dates.
    There are two sets of treatment-control comparisons: one for new hires, and another for incumbents.

    \subsubsection*{New Hires at Treated VS Control Firms (for Parent Shares)}

        Treated firms in this analysis are newly compliant establishments introducing onsite childcare between 2016 and 2018 (in response to non-compliance penalties), while control firms are early compliers introducing such services prior to 2013.
        \autoref{tab:childcare_compare} shows that treated and control firms are similar in observable characteristics, which supports the credibility of this comparison.
        We use the following regression to compare changes in the share of working parents among new hires:
        \begin{equation} \label{eq:newhire_regression}
            Y_{jt} = \sum_k \delta_k D_k + \lambda_j + \tau_t + \epsilon_{jt}
        \end{equation}

        \noindent Observations are firms $j$ at date $t$ (weighted by the size of new hire cohorts in 2015). Outcome $Y_{jt}$ is the share of new hires with young children (aged 0-6) in date $t$ at firm $j$, and $\epsilon_{jt}$ is the error term.
        Covariates in this regression are establishment fixed effects $\lambda_j$, calendar time fixed effects $\tau_t$, and event-time indicators $D_k$.
        Subscript $k$ indicates periods relative to the introduction of onsite childcare, and event-time indicators $D_k$ always equal zero for control firms.

        The causal parameter of interest is $\delta_k$, the difference in parent shares between treated and control firms after controlling for calendar time trends and baseline differences across establishments.
        ``Parallel trends'' is the key identifying assumption, which states that parent shares $Y_{jt}$ for treated firms would have evolved according to calendar time trends $\tau_t$ without onsite childcare. In this specification, $\delta_k$ is the additional entry of parents attributed to onsite childcare.
        Parallel trends is a reasonable assumption, but it is not innocuous. Its validity relies on stabilizing treatment effects for early compliers, so that changes in parent shares reflect general time trends rather than ongoing effects of onsite childcare.
        We provide supporting evidence for this assumption by confirming that parent shares do not differ between treated and control firms (beyond baseline differences) prior to the introduction of onsite childcare ($\delta_k=0$ for $k < 0$).

        There are alternative identification strategies that we did not pursue for practical reasons.
        One is to use a regression discontinuity design to compare firms immediately above and below the policy eligibility threshold at 500 employees. Unfortunately, this strategy lacks statistical power as there are not enough firms in the vicinity of this threshold.
        Another alternative is a differences-in-differences comparison of eligible VS ineligible firms, or compliant VS non-compliant firms. In our examination of the data, firms differ on observable attributes and do not exhibit parallel trends for either of these comparisons. 

    \subsubsection*{Parent VS Non-Parent Employees within Treated Firms (for Wages and Separations)} 

        In the second analysis, we compare employees within newly compliant firms. Parents serve as the treated group, since they directly benefit from onsite childcare. Non-parents serve as the counterfactual control group, since they capture company-wide changes that also affect parents. We compare parents with non-parents in the following regression:
        \begin{equation} \label{eq:parent_regression}
            Y_{it} = \sum_k \delta_k D_k + \gamma_k (D_k \times P_i) + \rho P_i + \beta X_{it} + \epsilon_{it}
        \end{equation}
        \noindent Observations are workers $i$ at biannual date $t$, and $\epsilon_{jt}$ is the error term.
        Outcomes $Y_{it}$ are log salaries (annual earnings divided by the number of months worked during the year) and separations (= 1 if separated at date $t$, among those employed at date $t-1$).
        Covariates in this regression are time-varying controls (age, tenure, tenure squared) and interactions between event-time indicators $D_k$ and the parent indicator $P_i$.
        As before, event-time indicators $D_k$ equal 1 at $k$ periods relative to onsite childcare, which apply to both parents and non-parents since they belong to treated establishments.

        The causal parameter of interest is $\gamma_k$, the difference in outcomes between parents and non-parents.
        The parallel trends assumption is that changes for non-parents is a good representation of changes that parents would have experienced without onsite childcare.
        This is plausible since they both belong to the same establishments, and we provide supporting evidence for parallel trends prior to the event date ($\gamma_k=0$ for $k < 0$)

\subsection{Entries and Separations for Parents with Young Children} \label{subsec:childcare_results}
    
    Introducing onsite childcare attracts more entry by parents with young children, who benefit the most from these services.
    \autoref{fig:childcare_entry} plots event study coefficients for the impact of onsite childcare on the share of new hires with young children ($\delta_k$ in \autoref{eq:newhire_regression}).
    Each coefficient is the causal impact at $k$ months relative to onsite childcare, which is the difference between treated and control firms after controlling for calendar time trends and baseline differences across establishments.
    Positive coefficients for periods $k > 0$ indicate higher shares of parents among new hires after onsite childcare.
    Coefficients do not statistically differ from zero for $k < 0$, which supports the parallel trends assumption and confirms that onsite childcare does not affect outcomes prior to its introduction.

    Parents are not more likely to separate after onsite childcare, despite slower salary growth relative to non-parents.
    Each coefficient in \autoref{fig:childcare_separation} compares the probability of separating from the firm at $k$ months relative to onsite childcare ($\gamma_k$ in \autoref{eq:parent_regression}).
    Parents and non-parents do not exhibit differences in separation rates prior to onsite childcare, which remains similar even after its introduction.
    \autoref{fig:childcare_salary} compares log salaries for parents and non-parents within treated firms, which additionally controls for baseline differences in salaries (with person fixed effects) and income-specific time trends (income rank interacted with linear calendar time).
    Negative coefficients for periods $k > 0$ indicate lower salaries for parents after the introduction of onsite childcare, which occurs due to slower salary growth for parents relative to non-parents.
    Collectively, these figures suggest that parents are willing to absorb slower salary growth to stay at firms with onsite childcare.

\begin{figure}[ht]
    \centering



    \makebox[\textwidth][c]{\includegraphics[width=0.80\textwidth]{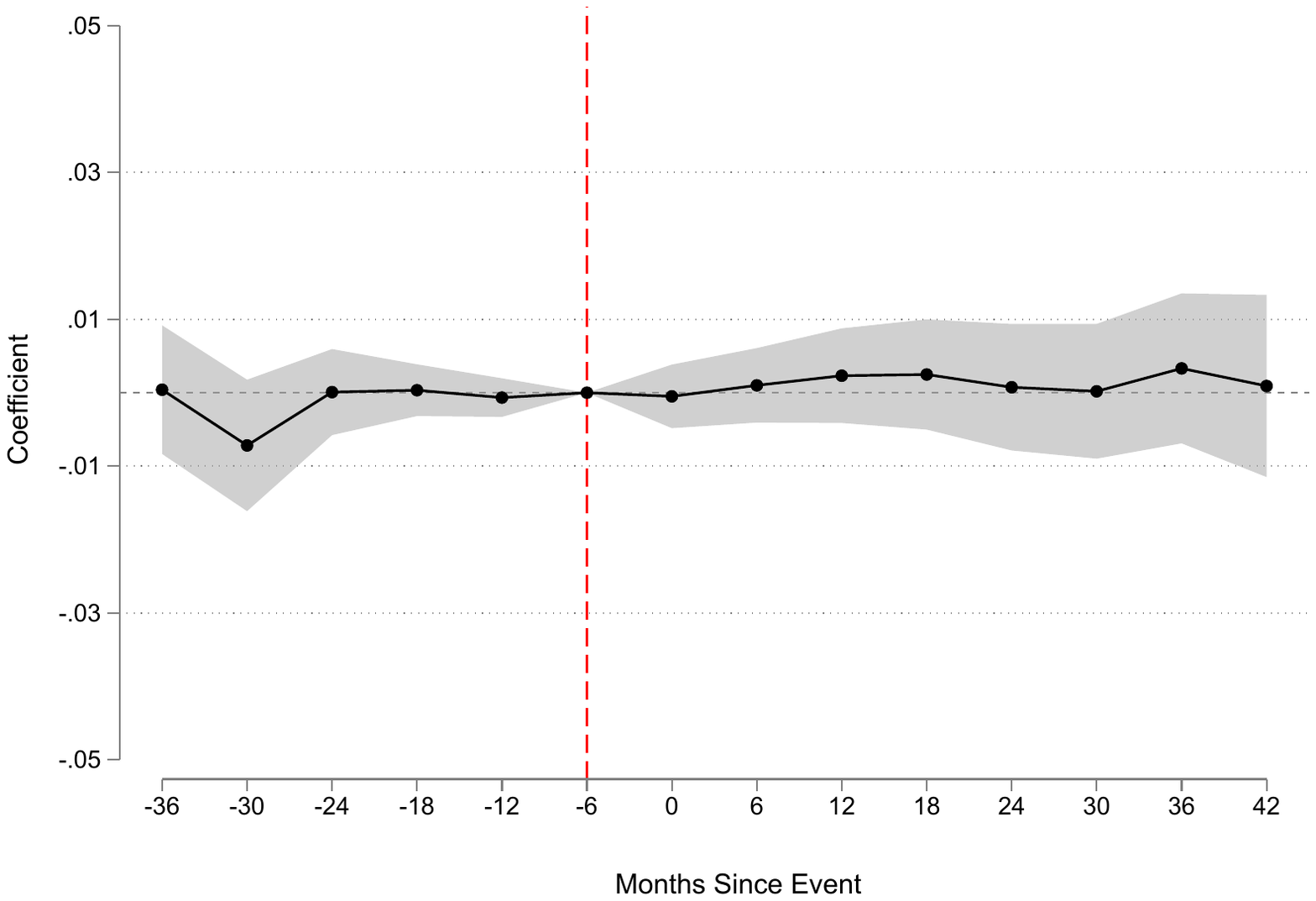}}

    \caption{Impact of Onsite Childcare on Log Salary (Parents of Young VS Older Children)} \label{fig:childcare_salary_older}
    \caption*{\normalsize 

        This figure plots event study coefficients for the impact of onsite childcare on log salaries for working parents (discussed in \autoref{subsec:childcare_results}). 
        The regression compares parents of younger children (age 0-6) with parents of older children (age 7+) within treated establishments installing onsite childcare facilities between 2016 and 2018.
        Observations are employees $i$ at date $t$.
        The outcome variable is log salary, which divides total earnings by the number of months worked during the year.
        The regression includes person fixed effects, time-varying attributes at the person level (age, tenure, tenure squared), linear time trends specific to each income rank, event-time indicators, and interactions between event-time indicators and an indicator for parents of young children.
        Event time (on the horizontal axis) is months relative to the installation of onsite childcare, and the indicator for $t=-6$ is omitted to avoid collinearity.
        Scatter points are coefficients on interaction terms, and shaded areas denote confidence intervals. 
        Each coefficient is the causal effect at event time $t$, interpreted as the difference between parents of younger and older children after controlling for person-level attributes and income-specific time trends.

        \, 

    }
\end{figure}

    Interestingly, \autoref{fig:childcare_salary_older} shows that parents with older children also exhibit similar changes in log salaries, even though they do not directly benefit from onsite childcare.
    This figure compares log salaries for parents with younger and older children within treated firms (ages 0-6 vs 7+), which exhibit no differences before and after the introduction.
    One interpretation is that onsite childcare only accelerates salary growth for childless employees, who enjoy competitive advantages with the rising share of newly hired parents who have less time available for work.
    Another possibility is that onsite childcare signals family-friendliness, and having more coworkers who are parents contributes to a family-friendly environment. This can appeal to parents of older children, who are willing to absorb slower salary growth to remain at such workplaces.


\section{Mandated Enrollment into Paternity Leave} \label{sec:patleave}

    The second quasi-experimental case study is mandated enrollment into paternity leave at a large conglomerate.
    Paid paternity leave was guaranteed by labor laws even before the mandate, but usage rates remain historically low. When the mandate was introduced, less than 5\% of fathers enrolled in paternity leave, although usage rates are steadily increasing over time.\footnote{ 
        Enrollment rates within two years of childbirth for fathers at control firms (subsidiaries of other conglomerates). This is likely an overstatement of national enrollment rates, which tends to be higher at conglomerates than smaller firms.
    } 
    Both surveys and anecdotes suggest that peer pressure makes it difficult for fathers to enroll in paternity leave, so an interesting aspect about this case is that the mandate significantly lowers barriers to enrollment. 

\subsection{Background on Mandated Paternity Leave} \label{subsec:patleave_background}

    We discuss key background and context behind the empirical analysis, but we cannot share specific details to avoid revealing the identity of the conglomerate. 
    This conglomerate mandated enrollment into paternity leave for all fathers employed in its subsidiaries, which we refer to as ``firms.''
    To our knowledge, the conglomerate did not introduce other family-friendly measures for men other than the mandate. 
    The mandate applies to regular employees with 6+ months of tenure, and it excludes part-time hires and temporary contractors.
    All fathers are required to enroll in at least one month of paternity leave within two years of childbirth.
    Fathers can extend this deadline by submitting waivers with explanatory statements, which must be approved by the human resources department.
    Fathers would enroll in paternity leave provided by the government (through employment insurance), and the conglomerate provides additional financial coverage to guarantee 100\% wage replacement.

    In principle, fathers can avoid this mandate by not reporting their childbirth. In practice, most fathers voluntarily report their childbirth to receive an unconditional ``congratulatory'' cash bonus.\footnote{ 
        At minimum, this bonus is 500k KRW for the first child and 2 million KRW for the second child ($\approx$ \$500 and \$2,000 in terms of purchasing power). Some subsidiaries offer a higher cash bonus.
    } 
    The conglomerate was proactive about enforcing this mandate among its subsidiaries, which received full support from the advisory board and entered into promotion criteria for managers.
    When a father does not enroll in paternity leave, the human resources department would contact both the employee and their manager to confirm their intended dates for enrollment.
    This mandate was heavily promoted both before and after its introduction, which applied to all fathers with childbirths after its enforcement.\footnote{There was no retroactive enforcement for childbirths preceding this announcement.} 
    More details on the conglomerate and the paternity leave mandate can be found in \textcite{tammy_jmp_2025}, which analyzes the same data and policy change to examine its impact on fertility decisions. 

\subsection{Differences-in-Differences (DiD) Design} \label{subsec:patleave_data_strategy}
 
    We use the same administrative data from \autoref{subsec:childcare_data}, but we additionally link the Parental Leave and Population Dynamic Registers to obtain month-year dates for childbirth and parental leave. Treated firms are subsidiaries in the conglomerate introducing the mandate, and control firms are subsidiaries in other conglomerates.
    \autoref{tab:patleave_compare} compares average characteristics for treated and control firms: treated firms are smaller in average revenue and employment, and 
    employees at treated firms are younger and have lower salaries. Treated employees are less likely to have children but slightly more likely to enroll in parental leave. We use the following regression to compare treated and control firms:
	\begin{equation} \label{eq:patleave_regression}
		Y_{it} = \lambda_{j(i)} + \sum_k \alpha_k D_{k(t)} + \beta_k (D_{k(t)} \times G_i) + \epsilon_{it}
	\end{equation}
    \noindent $\lambda_{j(i)}$ are firm fixed effects, $D_{k(t)}$ are relative time dummies ($k$ periods relative to the mandate), $G_i$ is an indicator for treated firms, and $\epsilon_{it}$ is the error term.
    The unit of observation depends on the outcome $Y_{it}$. For parental leave enrollment, observations are fathers $i$ with childbirths at date $t$. For parental shares among new hires, observations are new hire cohorts at date $t$ for firm $i$. For the share of staying parents, observations are firm $i$ at date $t$.

    The causal parameter of interest is $\beta_k$, the  difference in outcomes between treated and control firms at $k$ periods relative to the mandate.
    A causal interpretation of this parameter relies on the parallel trends assumption, which states that observed changes for control firms reflect how treated firms would have changed without the mandate.
    Since \autoref{tab:patleave_compare} shows that treated and control firms differ on observable characteristics, we take special care to provide supporting evidence for parallel trends prior to the mandate ($\beta_k = 0$ for $k < 0$).

\subsection{Increased Enrollment into Paternity Leave} \label{subsec:patleave_enroll}

    \autoref{fig:patleave_enroll} shows that mandating paternity leave was highly effective at increasing enrollment.
    Observations in this regression are fathers $i$ with childbirths at date $t$, and the outcome is enrollment into paternity leave within two years of childbirth (binary).
    This figure plots coefficients for the causal impact on enrollment ($\beta_k$ in \autoref{eq:patleave_regression}), which is the difference between fathers at treated and control firms after controlling for time trends and baseline differences across firms.
    Coefficients do not statistically differ from zero prior to the mandate, which offers supporting evidence for the parallel trends assumption.
    After the mandate, paternity leave enrollment sharply increases by around 60 percentage points.
    \textcite{tammy_jmp_2025} report similar results for the impact of the mandate on leave take-up among fathers.

    This increase in enrollment was mostly driven by fathers who complied with the mandate at the minimum duration (1 month). 
    Stacked area plots in \autoref{fig:patleave_duration} show the distribution of paternity leave durations for childbirth cohorts at treated and control firms.
    The pink area denotes non-enrollment, which does not necessarily imply non-compliance: some fathers are exempt from the mandate because they are part-time employees or temporary contractors.
    Non-pink regions denote enrollment at various durations, which gradually increases over time at control firms.
    At treated firms, there is a sharp increase in enrollment immediately after the mandate, which is mostly driven by fathers with exactly 1 month of paternity leave.
    While the mandate is effective at increasing enrollment on the extensive margin, there are no intensive margin effects as most fathers do not go beyond the required minimum.

\subsection{Entries and Separations for Fathers} \label{subsec:patleave_results}
    
    Mandated paternity leave attracted more entry by future fathers who would benefit from this change.
    \autoref{fig:patleave_entry} plots the impact on the share of new hires with impending childbirths, which we define as having children after their hire date.
    Each coefficient is the difference between treated and control firms after controlling for time trends and baseline differences across firms.
    There is no difference in this share before the mandate, which significantly increases for new hires after the mandate.
    This does not necessarily imply that newly hired fathers immediately benefit from the policy, but it does imply more entry by fathers who will eventually benefit from paternity leave during their tenure.
    On the other hand, the share of newly hired women with impending childbirths do not differ before and after the mandate, which is meaningful since the conglomerate expanded maternity leave by 12 months when they enforced the mandate.\footnote{
        This is 12 additional months of \underline{unpaid} leave on top of 12 months of paid leave provided through the government.
    } One interpretation is that working parents are more responsive to guaranteed usage of parental leave rather than duration expansions, although we cannot make definitive statements since mothers and fathers may react differently to policy changes.

    Fathers with young children were more likely to stay after the mandate, while men without children were more likely to leave.
    Each coefficient in \autoref{fig:patleave_retention} is the impact on the share of staying parents and non-parents.
    For these shares, the denominator is the number of men hired before the mandate, and the numerator is the number of fathers and childless men remaining at each date.
    Prior to the mandate, treated and control firms exhibit no differences in the propensity to stay and have children (in blue), which gradually expanded only after the mandate.
    These changes indicate that more parents are encouraged to stay after mandating paternity leave, which is possibly accompanied with family-friendly changes in workplace norms.
    On the other hand, men without children were more likely to leave after the mandate, although this difference is not statistically significant (in orange).
    One reason might be that these changes increase the burden of work on childless employees, who choose to pursue career opportunities elsewhere if they are not compensated with higher pay.\footnote{
        Many surveys and anecdotes indicate that increasing the burden of work on coworkers is a significant source of peer pressure against parental leave.
    }


\section{Family-Friendly Benefits and Labor Force Survival} \label{sec:separation}

    The common theme across quasi-experimental studies is that family-friendly benefits attract more entry by parents while reducing their separations.
    Which margin is more important for sorting on family-friendliness — job switching or separations?
    Using richer data on a wider range of benefits, we show that sorting on family-friendliness after childbirth mainly occurs through labor force survival rather than switching into new jobs.

\subsection{Data on Family-Friendly Benefits} \label{subsec:separation_data}

    We augment the administrative data with two crowdsourced datasets on workplace benefits.
    The first dataset comes from \href{https://bokziri.com/}{Bokziri.com}, a platform with crowdsourced information on workplace benefits offered at each company.
    There is a wide range of benefits on this platform, ranging from remote work, housing benefits, vacation policy, career development programs, onsite facilities, and family-related support (among many others).
    Details on workplace benefits are submitted via surveys to platform organizers, which are confirmed with human resource representatives at respective companies.

    The second dataset comes from \href{https://www.teamblind.com/kr/}{TeamBlind.com}, an anonymous discussion platform that features crowdsourced ratings on each company.
    Companies are rated on a 5-point Likert scale for five categories: work-life balance, career growth, compensation, management, and company culture.
    These ratings are comparable to Glassdoor.com, but TeamBlind possesses several key advantages.
    First, there is an additional layer of verification through corporate email addresses, which is required for all users on the platform.\footnote{
        Teamblind is offered on two platforms: one for Korean users, and another for English users. Corporate email verification is required for the Korean platform but optional for the English platform.
    }
    Second, it is predominantly used by job seekers in Korea. Job seekers regularly examine these ratings when making decisions on applications or job offers, and it is common for HR departments to monitor activity on this platform.
    As of March 2025, there are 12 million users on the platform that account for 90\% of employees in the top 1,000 firms (ranked by market value).\footnote{Source: \colorcite{kim2025teamblind, han2023teamblind}}

    \,

    \begin{samepage}
        \noindent Using this data, we mark employers as ``family-friendly'' if they offer any of the following benefits:
        \begin{enumerate}
            \setlength{\itemsep}{-4pt}
            \item[] \vspace{-2.5em}
            \item Work-life balance ratings above 4 (out of 5)
            \item Benefits regarding fertility, pregnancy, or childcare \\(e.g. onsite childcare facilities, or paid time off after childbirth/miscarriage)
            \item Family-friendly work arrangements \\(e.g. flexible work hours, remote/hybrid work, limits on late-night/overtime hours). 
        \end{enumerate}
    \end{samepage}

    \noindent Both platforms primarily cater to younger workers with higher salaries at large companies.
    We align the target population to this demographic by focusing on workers in 1983-95 birth cohorts who are employed at registered corporations with 50+ employees.
    We restrict the analysis to full-time workers earning annual salaries above the minimum wage but below 100 million KRW ($\approx$ \$100k in purchasing power), and we focus on workers with childbirths during the sample period.

    \autoref{tab:sumstat_jobspells} compares job spells in the target population and a subset with available data on workplace benefits through TeamBlind or Bokziri.
    Data on workplace benefits cover 53\% of job spells in the target population. On average, these jobs pay higher salaries, are more often held by men, and are slightly more likely to overlap with childbirth.
    Around half of these job spells are at family-friendly firms, and this share is lower for the full population as firms with missing data on workplace benefits are zero-coded for family-friendliness.

\subsection{Higher Survival Rates at Family-Friendly Firms} \label{subsec:separation_survival}

    \autoref{fig:survival_friendly} shows that parents are less likely to separate from firms with family-friendly benefits, which aligns with patterns observed in quasi-experimental studies.
    This figure plots Kaplan-Meier survival functions for parents employed at family-friendly firms (in blue) and other firms (in orange) at 15 months prior to childbirth.
    Both mothers and fathers are more likely to survive at family-friendly firms, but they differ in their trajectories of survival rates.
    Survival rates for fathers fall steadily over time, while mothers exhibit two periods with high separation hazards.
    The first is a steep decline in survival prior to childbirth $(t < 0)$, and the second is a sharp fall at the end of paid maternity leave $(t = 15)$.
    For both periods, separation hazards are larger for mothers at firms without family-friendly benefits.
    Although separations may arise from either job transitions or labor-force withdrawals, we later discuss evidence that they are largely driven by labor-force withdrawals for mothers.
    
    Differences across industries and amenities unrelated to family-friendliness cannot fully explain higher survival rates at family-friendly firms.
    \autoref{tab:hazard_regressions} reports coefficients and standard errors (in parentheses) from cox proportional hazard models for parents with childbirths during their tenure.
    Column 1 shows that separation hazards are lower at firms with family-friendly benefits.
    The magnitude of this sensitivity remains unchanged after controlling for company ratings on workplace culture, career growth, and management in columns 2-4 (on a 5-point Likert scale).
    Columns 5 and 6 show that controlling for log salaries and industry indicators reduces sensitivity to family-friendly benefits, which is expected to a certain extent.
    Firms and industries with better amenities tend to offer higher salaries \parencite{sockin_show_2024}, which makes parents less likely to separate.
    Including the full set of covariates in column 7 trims the coefficient on family-friendliness by a fourth of its magnitude, implying that these factors can only partially explain the sensitivity of separations to family-friendly benefits.

\subsection{Labor Force Withrawals from Firms Without Family-Friendly Benefits} \label{subsec:separation_share}

    Mothers at firms without family-friendly benefits are more likely to withdraw from the labor force after childbirth.
    The top panel of \autoref{fig:share_employed} plots employment rates for parents with childbirths between July 2017 and June 2018 (balanced panel with 10 quarters before and after childbirth).
    In this figure, employment is a broad definition that includes part-time and full-time jobs covered through the unemployment insurance agency (which excludes freelancers, gig workers, temporary contractors, and self-employed individuals).
    At $t = -10$ quarters before childbirth, employment rates are similar for mothers and fathers at 72\%.
    Fathers maintain similar employment rates after childbirth, but employment rates for mothers decline to 40\% after childbirth.
    To better understand where labor force withdrawals come from, bottom panels plot employment rates separately by family-friendliness at initial jobs.
    Among parents employed at $t = -10$ quarters before childbirth, left and right panels correspond to parents at firms with and without family-friendly benefits (respectively).
    Blue lines in both panels show that employment rates for fathers gradually converge to levels seen in the top panel regardless of family-friendliness at initial jobs.
    However, orange lines show that mothers are more likely to leave the labor force from firms without family-friendly benefits.
    At $t = 10$ quarters after childbirth, the employment rate is 60\% for mothers initially at family-friendly firms and only 40\% at firms without such benefits.

    Working mothers are more likely to be at family-friendly firms after childbirth, which is not driven by switching into new jobs.
    \autoref{fig:share_friendly} plots the share of working parents at family-friendly firms before and after childbirth.
    The left figure corresponds to an unbalanced panel of all parents who remain in the labor force.
    Mothers are 6 percentage points more likely to be at family-friendly firms after childbirth, while this share remains constant for fathers. 
    The figure on the right is for a balanced panel of parents who are always employed, which exhibits a constant share before and after childbirth.
    Among mothers employed at $t = -3$ quarters, only 2.6\% switch into new jobs after childbirth (at $t = 3$) while 43\% withdraw from the labor force.
    Taken together, these findings suggest that the rising share for mothers cannot be explained by switching into new jobs. Instead, mothers are concentrated at family-friendly firms because they exit the labor force after childbirth when their employers lack such benefits.


\section{Simple Model of Sorting After Childbirth} \label{sec:model}

    The previous section highlights three empirical facts about mothers at family-friendly firms: 1) lower separations, 2) higher labor force survival, and 3) a rising share of mothers at such firms after childbirth. We interpret these findings with a simple model of sorting after childbirth that features unobserved heterogeneity in outside options and opportunity costs for staying employed.

\subsection{Sorting Model with Unobserved Heterogeneity} \label{subsec:model_setup}

    Workers receive wage offers and choose between three options: 1) staying at their current jobs, 2) switching into new jobs, and 3) transitioning into non-employment. There is unobserved heterogeneity in the value of non-wage amenities $\epsilon_{if}$, non-employment $\eta_{if}$, and labor productivity $\psi_{if}$ across workers $i$ and firms $f$. The utility of accepting a job offer with wage $w$ and non-wage amenities $\epsilon_{if}$ is given by
    \refstepcounter{equation} \label{eq:utility_transition}
    \begin{equation}
        u(w, \epsilon_{if} \,|\, w_0) = (w-w_0) + \epsilon_{if}
        \tag*{(\theequation): Utility of Job Switching}
    \end{equation}
    $w_0$ is the current wage, $w-w_0$ is wage growth when accepting the offer, and $\epsilon_{if} \sim F_\epsilon(\cdot)$ is the relative value of non-wage amenities at the new job compared with the current job. 
    Since remaining at the current job involves no change in wages or amenities, the utility of staying at the current job is 0.
    
    Workers exit the labor force by transitioning into non-employment. They would lose their current wage $w_0$ by doing so, but they gain the relative value of non-employment $\eta_{if}$ (compared with their current job). The utility of transitioning into non-employment is given by: 
    \refstepcounter{equation} \label{eq:utility_exit}
    \begin{equation}
        u(\eta_{if} \,|\, w_0) = -w_0 + \eta_{if}
        \tag*{(\theequation): Utility of Non-Employment}
    \end{equation}
    \noindent $\eta_{if} \sim F_\eta(\cdot)$ is the value of non-employment, which reflects the opportunity cost of labor (e.g. hiring baby sitters, home maintenance, and less time available for children). 
    
    On the demand side, monopsonistic firms set wage offers $w$ to maximize profits. Firms face uncertainty in how candidates value non-wage amenities ($\epsilon_{if}$) and non-employment ($\eta_{if}$), which is private to the job seeker. Given this uncertainty, expected profits are given by:
    \refstepcounter{equation} \label{eq:expected_profits}
    \begin{align}
        \pi(w | w_0) = a(w \, | \, w_0) \cdot (\psi_{if}-w)
        \tag*{(\theequation): Expected Profits}
    \end{align}
    \noindent where $w$ is the offered wage, $w_0$ is the current wage, and $\psi_{if} \sim F_\psi(\cdot)$ is labor productivity known to the firm. 
    Firms' uncertainty is reflected in $a(w \, | \, w_0) = P(switch)$, the probability that the job seeker accepts wage offer $w$ given their current wage $w_0$. 
    Profit-maximizing wage offers $w^*$ are increasing in labor productivity $\psi_{if}$, so productive workers are more likely to pursue job transitions as they face better outside options with higher wages.

\subsection{Probability of Staying, Job Switching, and Non-Employment} \label{subsec:model_prob}
    
    Workers switch into new jobs if doing so gives them higher utility than both non-employment and staying at their current job. Similarly, workers transition into non-employment if doing so is better than their current job and offered job. Given heterogeneity in $\epsilon_{if}$ and $\eta_{if}$, choice probabilities are given by:\footnote{Note that these expressions assume $\epsilon_{if} \perp \eta_{if}$. This assumption can be relaxed with an integral over the joint density, which involves more algebra but retains the same intuition.}
    \begin{alignat*}{4}
        &  P(switch) &&= P\big(\epsilon_{if} > -(w-w_0) \big) &&\cdot P\big(\epsilon_{if} > -(w-\eta_{if}) \big) \; &&\longmapsto \; \text{Probability of Job Switching}  \\ 
        & P(exit) &&= P\big(\eta_{if} > w_0 \big) &&\cdot P\big(\eta_{if} > w + \epsilon_{if} \big) \; &&\longmapsto \; \text{Probability of Non-Employment} \\ 
        & P(stay) &&= 1-P(switch)-P(exit) && &&\longmapsto \; \text{Probability of Staying at Current Job}
    \end{alignat*}
    \noindent The left panel of \autoref{fig:choice_prob} shows choice probabilities for parents employed at $t = -10$ quarters before childbirth. Bars indicate proportions of parents who stay at their jobs (``stay''), switch into new jobs (``switch''), and enter into non-employment (``exit''), based on their employment status at $t = -4$ quarters \underline{before} childbirth. The right panel is the corresponding figure for parents employed at $t = -3$ quarters before childbirth, with bars indicating choice probabilities \underline{after} childbirth (at $t = 3$). Orange and blue bars correspond to mothers and fathers, while thicker and thinner colors correspond to firms with and without family-friendly benefits (respectively).

\subsection{Changes in Unobserved Heterogeneity After Childbirth} \label{subsec:model_changes}

    \begin{samepage}
    The model can intuitively explain key empirical facts with changes in unobserved heterogeneity after childbirth, which shape these choice probabilities. 
    Let $(\mu_\epsilon$, $\mu_\eta$, $\mu_\psi)$ denote average values for non-wage amenities, non-employment, and labor productivity (respectively). While there are many plausible changes in $(\mu_\epsilon$, $\mu_\eta$, $\mu_\psi)$ after childbirth, we allow them to differ along three attributes: 
        \begin{itemize}
            \setlength{\itemsep}{-4pt}
            \item \textit{Gender}: $m_1$ for men, $m_0$ for women
            \item \textit{Childbirth Status}: $c_1$ with children, $c_0$ without children
            \item \textit{Family-Friendliness}: $f_1$ for family-friendly firms, $f_0$ for other firms
        \end{itemize}
    \end{samepage}
    
\begin{figure}[ht]

    \makebox[\textwidth][c]{
        \includegraphics[width=0.45\paperwidth]{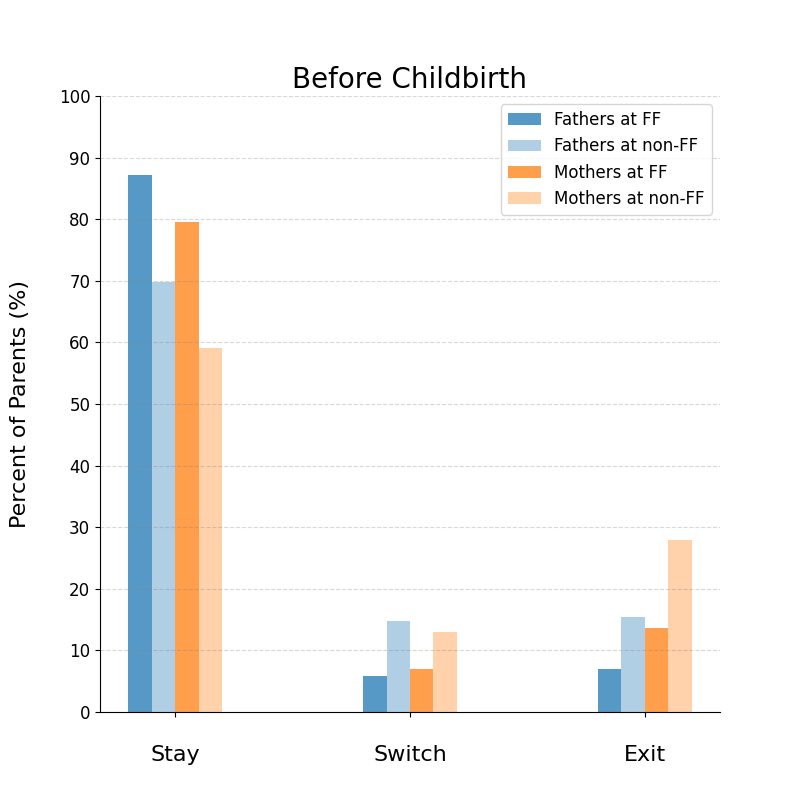}
        \includegraphics[width=0.45\paperwidth]{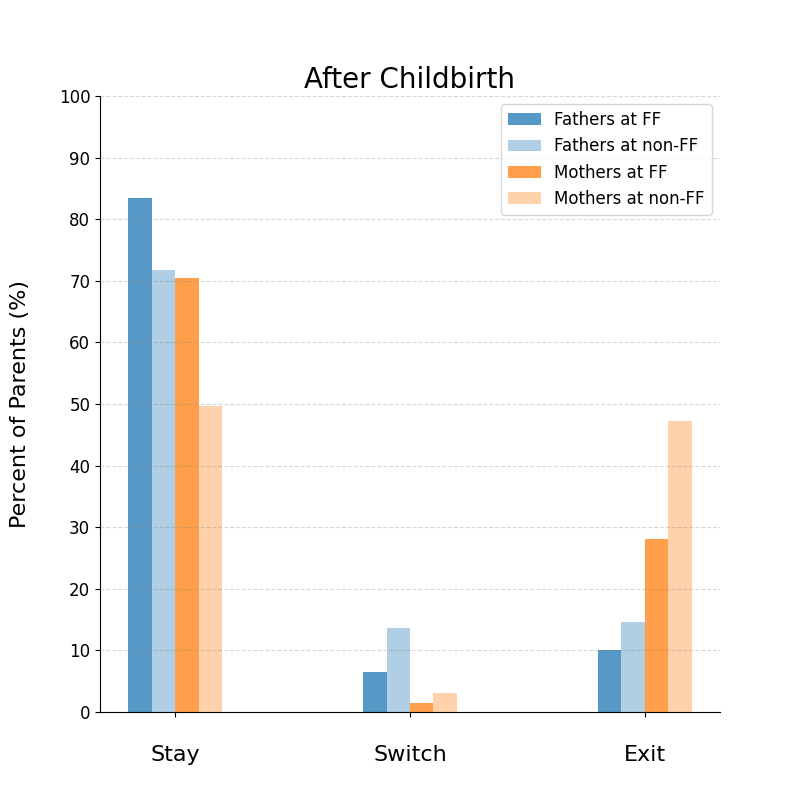}
    }

    \caption{Probability of Staying, Job Switching, and Non-Employment} \label{fig:choice_prob}
    \caption*{\normalsize 
        Bars indicate the probability of staying at the initial job (``stay''), switching into a new job (``switch''), or entering into non-employment (``exit''), which we discuss in \autoref{subsec:model_prob}. 
        The left panel is for parents employed 10 quarters prior to childbirth, and bars indicate the proportion in each state six quarters later (at 4 quarters \underline{before} childbirth). The right panel is for parents employed 3 quarters prior to childbirth, and their bars indicate proportions in each state six quarters later (at 3 quarters \underline{after} childbirth).

        \,

        Orange and blue bars correspond to mothers and fathers (respectively). Thicker and ligher bars corresponds to firms with and without family-friendly benefits (respectively). Bars sum to 100\% for each combination of color and thickness (e.g. thick blue bars for fathers at family-friendly firms before childbirth). The sample consists of parents in 1983-95 birth cohorts with childbirths between July 2017 and June 2018. Employment includes part-time and full-time jobs at any firm with unemployment insurance (UI) coverage, and family-friendliness is zero-coded for missing data on workplace benefits. 
    }
\end{figure}

    \noindent First, consider higher labor force survival for mothers at family-friendly firms in \autoref{fig:share_employed}.
    In the right panel of \autoref{fig:choice_prob}, $P(exit)$ is 19 percentage points lower for mothers at such firms (thicker - thinner orange bars).
    This can be explained with lower value of non-employment $(\eta)$ for mothers $(m_0)$ at family-friendly firms $(f_1)$, which corresponds to $\mu_\eta(m_0, f_1) < \mu_\eta(m_0, f_0)$ in the model.
    Family-friendly benefits lower the opportunity cost of staying employed since it becomes easier for mothers to meet parenting needs for their children (e.g. onsite childcare, flexible hours for daycare pickups, remote work while breastfeeding).

    Second, recall that separation rates are lower at firms with family-friendly benefits in \autoref{fig:survival_friendly}.
    For mothers, these differences are entirely due to labor force exits since their $P(switch)$ is close to zero after childbirth.\footnote{Orange bars for $P(switch)$ in the right panel of \autoref{fig:choice_prob}}
    Fathers continue to switch jobs after childbirth,\footnote{In \autoref{fig:choice_prob}, blue bars for $P(switch)$ are similar for the left and right panels.}
    but they are less likely to separate from family-friendly firms since it is harder to find better amenities elsewhere.
    This corresponds to $\mu_\epsilon(f_1) < \mu_\epsilon(f_0)$ in the model, which implies lower $P(switch)$ and higher $P(stay)$ at family-friendly firms.

    Finally, consider the rising share of mothers at family-friendly firms in \autoref{fig:share_friendly}.
    This is primarily driven by high $P(exit)$ for mothers at firms without family-friendly benefits, since there are higher opportunity costs $\eta$ for staying employed during infant care.
    Why doesn't \autoref{fig:choice_prob} indicate higher $P(switch)$ for mothers after childbirth, especially when family-friendly benefits become more important to them?
    Most mothers take on a heavier role during infant care (e.g. breastfeeding), which makes them less productive after childbirth. 
    Mothers have less time available for work, and they have even less time to prepare for interviews and actively pursue job transitions.
    This corresponds to $\mu_\phi(m_0, c_1) < \mu_\phi(m_0, c_0)$ in the model, which lowers $P(switch)$ due to worse outside options with lower wages.
    While taking care of newborns, the feasible option for most mothers is to either keep their current job or stop working altogether.


\section{Policy Implications} \label{sec:conclusion}

This paper analyzed patterns of separations and job transitions to study how parents sort into family-friendly firms. 
We examined two quasi-experimental cases: 1) staggered compliance with onsite childcare, and 2) mandated paternity leave at a large conglomerate. 
Both cases reveal that introducing family-friendly changes attract more entry by parents who would gain from these benefits, 
and parents with young children stay despite slower salary growth. 
When using data on a wider range of benefits, we find that mothers are concentrated at family-friendly firms because they exit the labor force after childbirth when their employers lack such benefits. 
Our model of sorting explains these results with heterogeneity in outside options and opportunity costs for staying employed, 
which change after childbirth and vary by gender and family-friendliness at current jobs.

What are the policy implications of these findings? 
In developed countries with rapidly declining fertility like Korea, it is more important than ever to encourage mothers to remain in the labor force to reduce strains on welfare programs like social security or public health insurance. 
Most mothers with newborns do not have the capacity to actively switch into family-friendly firms, as it is already difficult to make time available for their current jobs. 
Mothers with newborns are more likely to leave the labor force than transition into family-friendly jobs, and mothers who continue to work are tied to their employers shortly after childbirth.

Family-friendly benefits make it easier for women to balance childcare with workplace demands after childbirth.
Prior studies show that universal childcare is effective at encouraging female labor force participation, especially in countries with low baseline participation rates (see \colorcite{cortes_children_2023} for a thorough review).
Our findings explain why this might be the case: mothers with newborns are unable to switch into new jobs that allow them to better balance family with work, so better access to family-friendly benefits prevents labor force leakage from firms without such benefits.

Since labor force withdrawals mostly occur at firms without family-friendly benefits, governments should target resources to this segment of the labor market.
For example, subsidies for family-friendly benefits can be targeted to smaller firms that cannot cost-effectively provide these benefits at scale. 
Alternatively, governments can step in to coordinate these benefits for smaller firms while mandating their provision for larger firms (e.g. mandating onsite childcare for larger firms while expanding public childcare for smaller firms).
Encouraging mothers to remain in the labor force would be an effective way to broaden the tax base in an era of declining fertility and fiscal pressure on social welfare programs.


\clearpage
\addcontentsline{toc}{section}{Supporting Materials}
\addcontentsline{toc}{subsection}{References}
\printbibliography[title=References]

@online{kim2025teamblind,
  author    = {Kim, Mun Sun},
  title     = {직장인 소셜 플랫폼 블라인드, 가입자 1,200만 명 돌파 [Social Platform TeamBlind Surpasses 12 Million Users]},
  year      = {2025},
  url       = {https://platum.kr/archives/254551},
  urldate   = {2025-05-08},
  langid    = {korean},
}

@online{han2023teamblind,
  author    = {Han, Dong Hyun},
  title     = {대기업 10명 중 8명은 블라인드 사용… 가입자 수 800만 돌파 [8 out of 10 Employees in Conglomerates are on TeamBlind, which Surpasses 8 Million Users]},
  year      = {2023},
  url       = {https://www.seoulwire.com/news/articleView.html?idxno=495354},
  urldate   = {2025-05-08},
  langid    = {korean},
}

@article{cortes_children_2023,
	title = {Children and the Remaining Gender Gaps in the Labor Market},
	volume = {61},
	issn = {0022-0515},
	url = {https://pubs.aeaweb.org/doi/10.1257/jel.20221549},
	doi = {10.1257/jel.20221549},
	pages = {1359--1409},
	number = {4},
	journaltitle = {Journal of Economic Literature},
	shortjournal = {Journal of Economic Literature},
	author = {Cortés, Patricia and Pan, Jessica},
	urldate = {2025-03-18},
	date = {2023-12-01},
	langid = {english},
}

@article{chernozhukov_doubledebiased_2018,
	title = {Double/debiased machine learning for treatment and structural parameters},
	volume = {21},
	rights = {http://doi.wiley.com/10.1002/tdm\_license\_1.1},
	issn = {1368-4221, 1368-423X},
	url = {https://academic.oup.com/ectj/article/21/1/C1/5056401},
	doi = {10.1111/ectj.12097},
	pages = {C1--C68},
	number = {1},
	journaltitle = {The Econometrics Journal},
	author = {Chernozhukov, Victor and Chetverikov, Denis and Demirer, Mert and Duflo, Esther and Hansen, Christian and Newey, Whitney and Robins, James},
	urldate = {2025-05-10},
	date = {2018-02-01},
	langid = {english},
}

@article{kleven_child_2019,
	title = {Child Penalties across Countries: Evidence and Explanations},
	volume = {109},
	issn = {2574-0768, 2574-0776},
	url = {https://pubs.aeaweb.org/doi/10.1257/pandp.20191078},
	doi = {10.1257/pandp.20191078},
	shorttitle = {Child Penalties across Countries},
	pages = {122--126},
	journaltitle = {{AEA} Papers and Proceedings},
	shortjournal = {{AEA} Papers and Proceedings},
	author = {Kleven, Henrik and Landais, Camille and Posch, Johanna and Steinhauer, Andreas and Zweimüller, Josef},
	urldate = {2025-05-10},
	date = {2019-05-01},
	langid = {english},
}

@article{kleven_children_2019,
	title = {Children and Gender Inequality: Evidence from Denmark},
	volume = {11},
	issn = {1945-7782, 1945-7790},
	url = {https://pubs.aeaweb.org/doi/10.1257/app.20180010},
	doi = {10.1257/app.20180010},
	shorttitle = {Children and Gender Inequality},
	pages = {181--209},
	number = {4},
	journaltitle = {American Economic Journal: Applied Economics},
	shortjournal = {American Economic Journal: Applied Economics},
	author = {Kleven, Henrik and Landais, Camille and Søgaard, Jakob Egholt},
	urldate = {2025-05-10},
	date = {2019-10-01},
	langid = {english},
}

@article{kleven_child_2025,
	title = {The Child Penalty Atlas},
	journaltitle = {Review of Economic Studies},
	author = {Kleven, Henrik and Landais, Camille and Leite-Mariante, Gabriel},
	date = {2025},
	langid = {english},
}

@article{sockin_show_2024,
	title = {Show Me the Amenity: Are Higher-Paying Firms Better All Around?},
	journaltitle = {{CESifo} Working Paper 9842},
	author = {Sockin, Jason},
	date = {2024},
	langid = {english},
}

@article{kuziemko_mommy_2018,
	title = {The Mommy Effect: Do Women Anticipate the Employment Effects of Motherhood?},
	url = {http://www.nber.org/papers/w24740.pdf},
	shorttitle = {The Mommy Effect},
	pages = {w24740},
	journaltitle = {{NBER} Working Paper No. 24740},
	author = {Kuziemko, Ilyana and Pan, Jessica and Shen, Jenny and Washington, Ebonya},
	urldate = {2025-06-02},
	date = {2018-06},
	langid = {english},
	doi = {10.3386/w24740},
}

@article{cha_overwork_2013,
	title = {Overwork and the Persistence of Gender Segregation in Occupations},
	volume = {27},
	issn = {0891-2432, 1552-3977},
	url = {https://journals.sagepub.com/doi/10.1177/0891243212470510},
	doi = {10.1177/0891243212470510},
	pages = {158--184},
	number = {2},
	journaltitle = {Gender \& Society},
	shortjournal = {Gender \& Society},
	author = {Cha, Youngjoo},
	urldate = {2025-06-30},
	date = {2013-04},
	langid = {english},
}

@article{corradini_collective_2025,
	title = {Collective Bargaining for Women: How Unions Can Create Female-Friendly Jobs},
	issn = {1556-5068},
	url = {https://www.ssrn.com/abstract=4219409},
	doi = {10.2139/ssrn.4219409},
	shorttitle = {Collective Bargaining for Women},
	journaltitle = {Quarterly Journal of Economics},
	shortjournal = {{SSRN} Journal},
	author = {Corradini, Viola and Lagos, Lorenzo and Sharma, Garima},
	urldate = {2025-06-30},
	date = {2025},
	langid = {english},
}

@article{maestas_value_2023,
	title = {The Value of Working Conditions in the United States and Implications for the Structure of Wages},
	volume = {113},
	issn = {0002-8282},
	url = {https://pubs.aeaweb.org/doi/10.1257/aer.20190846},
	doi = {10.1257/aer.20190846},
	pages = {2007--2047},
	number = {7},
	journaltitle = {American Economic Review},
	shortjournal = {American Economic Review},
	author = {Maestas, Nicole and Mullen, Kathleen J. and Powell, David and Von Wachter, Till and Wenger, Jeffrey B.},
	urldate = {2025-06-30},
	date = {2023-07-01},
	langid = {english},
}

@article{mas_valuing_2017,
	title = {Valuing Alternative Work Arrangements},
	volume = {107},
	issn = {0002-8282},
	url = {https://pubs.aeaweb.org/doi/10.1257/aer.20161500},
	doi = {10.1257/aer.20161500},
	pages = {3722--3759},
	number = {12},
	journaltitle = {American Economic Review},
	shortjournal = {American Economic Review},
	author = {Mas, Alexandre and Pallais, Amanda},
	urldate = {2025-06-30},
	date = {2017-12-01},
	langid = {english},
}

@article{pertold-gebicka_employment_2016,
	title = {Employment Adjustments Around Childbirth},
	issn = {1556-5068},
	url = {https://www.ssrn.com/abstract=2725046},
	doi = {10.2139/ssrn.2725046},
	journaltitle = {{IZA} Discussion Paper No. 9685},
	shortjournal = {{SSRN} Journal},
	author = {Pertold-Gebicka, Barbara and Pertold, Filip and Datta Gupta, Nabanita},
	urldate = {2025-06-30},
	date = {2016},
	langid = {english},
}

@article{wasserman_hours_2023,
	title = {Hours Constraints, Occupational Choice, and Gender: Evidence from Medical Residents},
	volume = {90},
	rights = {https://academic.oup.com/pages/standard-publication-reuse-rights},
	issn = {0034-6527, 1467-937X},
	url = {https://academic.oup.com/restud/article/90/3/1535/6648111},
	doi = {10.1093/restud/rdac042},
	shorttitle = {Hours Constraints, Occupational Choice, and Gender},
	pages = {1535--1568},
	number = {3},
	journaltitle = {The Review of Economic Studies},
	author = {Wasserman, Melanie},
	urldate = {2025-06-30},
	date = {2023-05-05},
	langid = {english},
}

@article{wiswall_preference_2018,
	title = {Preference for the Workplace, Investment in Human Capital, and Gender*},
	volume = {133},
	issn = {0033-5533, 1531-4650},
	url = {https://academic.oup.com/qje/article/133/1/457/4095201},
	doi = {10.1093/qje/qjx035},
	pages = {457--507},
	number = {1},
	journaltitle = {The Quarterly Journal of Economics},
	author = {Wiswall, Matthew and Zafar, Basit},
	urldate = {2025-06-30},
	date = {2018-02-01},
	langid = {english},
}

@article{hotz_parenthood_2018,
	title = {Parenthood, Family Friendly Workplaces, and the Gender Gaps in Early Work Careers},
	url = {http://www.nber.org/papers/w24173.pdf},
	pages = {w24173},
	journaltitle = {{NBER} Working Paper No. 24173},
	author = {Hotz, V. Joseph and Johansson, Per and Karimi, Arizo},
	urldate = {2025-06-30},
	date = {2018},
	langid = {english},
	doi = {10.3386/w24173},
}

@article{lachowska_work_2023,
	title = {Work Hours Mismatch},
	url = {http://www.nber.org/papers/w31205.pdf},
	pages = {w31205},
	journaltitle = {{NBER} Working Paper No. 31205},
	author = {Lachowska, Marta and Mas, Alexandre and Saggio, Raffaele and Woodbury, Stephen},
	urldate = {2025-06-30},
	date = {2023-05},
	langid = {english},
	doi = {10.3386/w31205},
}

@article{mas_non-wage_2025,
	title = {Non-Wage Amenities},
	url = {https://www.ssrn.com/abstract=5215044},
	doi = {10.2139/ssrn.5215044},
	journaltitle = {{NBER} Working Paper No. 33643},
	author = {Mas, Alexandre},
	urldate = {2025-05-20},
	date = {2025},
}

@article{tammy_jmp_2025,
	title = {Can Paternity Leave Boost Fertility?
Evidence from a Corporate Mandate in South Korea},
	journaltitle = {University of Michigan, Job Market Paper},
	author = {Lee T., Tammy and Lee J., Jungmin},
	date = {2025},
}

\clearpage

\section*{Main Figures}
\addcontentsline{toc}{subsection}{Main Figures}

\vspace*{\fill}
\vspace*{\fill}
\clearpage

\newgeometry{top=0.1in, bottom=0.8in, left=0.8in, right=0.8in}

\vspace*{\fill}
\vspace*{\fill}
\clearpage

\vspace*{\fill}
\vspace*{\fill}
\clearpage

\vspace*{\fill}
\vspace*{\fill}
\clearpage

\clearpage

\vspace*{\fill}
\vspace*{\fill}
\clearpage

\vspace*{\fill}
\vspace*{\fill}
\clearpage

\vspace*{\fill}
\vspace*{\fill}
\clearpage

\vspace*{\fill}
\vspace*{\fill}
\clearpage

\clearpage

\clearpage

\vspace*{\fill}
\vspace*{\fill}
\clearpage

\vspace*{\fill}
\vspace*{\fill}
\clearpage

\restoregeometry

\clearpage

\section*{Main Tables}
\addcontentsline{toc}{subsection}{Main Tables}

\vspace*{\fill}
\begin{table}[ht]
    \large
    \caption{Comparison of Treated and Control Establishments (Onsite Childcare)} \label{tab:childcare_compare}
    \begingroup
    \renewcommand{\arraystretch}{1.3}
    \makebox[\textwidth][c]{{
	\def\sym#1{\ifmmode^{#1}\else\(^{#1}\)\fi}
\begin{tabular}{l|cc|c}
\toprule[2pt]
                              &    Late Compliers       &   Early Compliers        & Difference    \\
                              &    (Treated) & (Control) &   (p-value)   \\
\hline
\multirow{2}{*}{Age} &         37.3 &         37.7 &       -0.413 \\
                              & (4.19)       & (3.97)       &       (0.49) \\
\multirow{2}{*}{Female} &         0.42 &         0.45 &       -0.030 \\
                        &         (0.14) &         (0.13) &       (0.88) \\
\multirow{2}{*}{Monthly Income (10k KRW)} &          505 &          527 &      -22.003 \\
                              &        (165) &        (169) &       (0.37) \\
\multirow{2}{*}{Average Tenure as of 2015 (years)} &        7.28 &         7.77 &       -0.489 \\
                        &       (4.43) &       (4.22) &       (0.44) \\
Parental Leave Usage (in 2015) &         0.10 &        0.09 &         0.010 \\
 &       (0.08) &      (0.07) &         (0.31) \\
\multirow{2}{*}{Full-Time Employees}  &        0.95 &        0.96 &     -0.008 \\
                              & (0.063) & (0.061) &  (0.39) \\
\multirow{2}{*}{Employment Size}  &        \multirow{2}{*}{1,846} &        \multirow{2}{*}{1,619} &       \multirow{2}{*}{226.410} \\
& & & \\ \hline
Number of Establishments &           96 &           93 &              \\
              
\bottomrule[2pt]
\end{tabular}
}}
    \endgroup

    \vspace{24pt}
    \caption*{\normalsize 
        Treated establishments are ``late adopters'' installing onsite childcare facilities between 2016 and 2018, while control establishments are ``early adopters'' installing such facilities before 2013 (discussed in \autoref{subsec:childcare_strategy}). Values inside parentheses are standard errors for averages and p-values for differences. Parental leave usage correspond to eligible parents with children aged 0-8.

        \,

    }
\end{table}
\vspace*{\fill}
\clearpage

\newgeometry{top=0.1in, bottom=0.8in, left=0.8in, right=0.8in}

\vspace*{\fill}
\begin{table}[ht]
    \large
    \caption{Comparison of Treated and Control Establishments (Mandated Paternity Leave)} \label{tab:patleave_compare}
    \begingroup
    \renewcommand{\arraystretch}{1.3}
    \makebox[\textwidth][c]{{
	\def\sym#1{\ifmmode^{#1}\else\(^{#1}\)\fi}

\begin{tabular}{l|cc|c}
    \toprule[2pt]
    & Treated & Control & Difference \\
    & (Average) & (Average) &  \\ \hline
    Age & 36.1 & 38.2 & -2.1$^{***}$ \\
    Age (Men) & 34.2 & 38.8 & -4.6$^{**}$ \\
    Monthly Salary in 10k KRW (Age 25-40) & 444.9 & 579.1 & -134.2$^{***}$ \\
    Monthly Salary in 10k KRW (Age 25-40, Men) & 471.9 & 616.2 & -144.3$^{***}$ \\ \hline
    \textit{Have Young Children (0-2 y.o.)} & & & \\ 
    \hspace{0.25em} Men & 0.07 & 0.11 & -0.04$^{***}$ \\
    \hspace{0.25em} Women & 0.03 & 0.04 & -0.01$^{***}$ \\ \hline
    \textit{Enrolled in Parental Leave (within 1 year)} & & & \\
    \hspace{0.25em} Men & 0.03 & 0.02 & 0.009 \\
    \hspace{0.25em} Women & 0.88 & 0.85 & 0.03 \\ \hline
    Employment Size & 711 & 823 & -80 \\
    Revenue (1 billion KRW) & 475.7 & 738.4 & -233 \\ \hline
    N Firms & 126 & 1936 & \\ 
   \bottomrule[2pt]
    \end{tabular}
}
}
    \endgroup

    \vspace{24pt}
    \caption*{\normalsize 
        Treated establishments belong to the conglomerate that mandated enrollment into paternity leave, and control establishments belong to other conglomerates that did not implement such policies (discussed in \autoref{subsec:patleave_data_strategy}). Averages correspond to values in the year prior to mandated paternity leave. *, **, and *** denote statistical significance at 10\%, 5\%, 1\% levels (respectively).

        \, 

    }
\end{table}
\vspace*{\fill}
\clearpage

\vspace*{\fill}
\begin{table}[ht]
    \large
    \caption{Summary Statistics for Job Spells} \label{tab:sumstat_jobspells}
    \begingroup
    \renewcommand{\arraystretch}{1.3}
    \makebox[\textwidth][c]{\centering
\begin{tabular}{l|cc|cc|cc}
\toprule[2pt]
\multicolumn{1}{c}{} & \multicolumn{2}{c}{\makecell[c]{Target Population}} & \multicolumn{2}{c}{\makecell[c]{Available Data \\on Benefits}} & \multicolumn{2}{c}{} \\
\cmidrule(r){2-3} \cmidrule(r){4-5}
\multicolumn{1}{c}{} & Mean & \multicolumn{1}{c}{Std Dev} & Mean & \multicolumn{1}{c}{Std Dev} & Diff & Std Err \\
\midrule
Age at Hire (years)                 & 29.6  & 3.6  & 29.7  & 3.6        & 0.08     & (0.01) \\
Starting Salary (10k KRW)           & 3270.3 & 1986.4 & 3497.9 & 2320.6 & 227.6   & (5.1) \\
Female (\%)                         & 0.419 & 0.493 & 0.368 & 0.482     & -0.051   & (0.001) \\ 
Family-Friendly Firm (\%)           & 0.264 & 0.441 & 0.498 & 0.500     & 0.234   & (0.001) \\ \hline
Married Before Hire (\%)            & 0.255 & 0.436 & 0.276 & 0.447     & 0.021   & (0.001) \\
Married During Tenure (\%)          & 0.267 & 0.442 & 0.248 & 0.432     & -0.020   & (0.001) \\ 
Married After Tenure (\%)           & 0.478 & 0.500 & 0.476 & 0.499     & -0.002   & (0.001) \\ \hline
Childbirth Before Hire (\%)         & 0.493 & 0.500 & 0.479 & 0.500     & -0.014   & (0.001) \\
Childbirth During Tenure (\%)       & 0.185 & 0.388 & 0.204 & 0.403     & 0.019   & (0.001) \\ 
Childbirth After Tenure (\%)        & 0.322 & 0.467 & 0.316 & 0.465     & -0.005   & (0.001) \\ \hline
Number of Job Spells & \multicolumn{2}{c|}{N = 575,555} & \multicolumn{2}{c|}{N = 304,533} & & \\
\bottomrule[2pt]
\end{tabular}}
    \endgroup

    \vspace{24pt}
    \caption*{\normalsize 
        This table compares average characteristics for job spells with and without available data on workplace benefits (discussed in \autoref{subsec:separation_data}). 
        Each observation is a job spell with hire dates between 2015 and 2020, which records tenure at separation (but includes non-terminated spells). 
        All job spells belong to workers who eventually become parents. 
        The target population is workers in 1983-95 birth cohorts who are employed at registered corporations with 50+ employees.
        We restrict the analysis to full-time employees earning annual salaries above the minimum wage but below 100 million KRW, and we focus on workers with childbirths during the sample period.
        For a subset of job spells, data on workplace benefits are available through TeamBlind or Bokziri.
        Employers are marked as ``family-friendly'' if they offer any of the following benefits:
        1) work-life balance ratings above 4 (out of 5), 
        2) benefits regarding childbirth, pregnancy, or childrearing (e.g. onsite childcare or paid time off after childbirth or miscarriage), or 
        3) family-friendly work arrangements (e.g. flexible work hours, remote/hybrid work, limits on late-night/overtime hours). 
        For the target population, firms with missing data on workplace benefits are zero-coded for family-friendliness. 
    }
\end{table}
\vspace*{\fill}
\clearpage

\vspace*{\fill}
\begin{table}[ht]
    \normalsize
    \caption{Sensitivity of Separation Hazards to Family-Friendly Benefits} \label{tab:hazard_regressions}
    \begingroup
    \renewcommand{\arraystretch}{1.3}
    \makebox[\textwidth][c]{{
\def\sym#1{\ifmmode^{#1}\else\(^{#1}\)\fi}
\begin{tabular}{l*{7}{c}}
\toprule[2pt]
                              &\multicolumn{1}{c}{(1)}   &\multicolumn{1}{c}{(2)}   &\multicolumn{1}{c}{(3)}   &\multicolumn{1}{c}{(4)}   &\multicolumn{1}{c}{(5)}   &\multicolumn{1}{c}{(6)}   &\multicolumn{1}{c}{(7)}   \\
\hline
Family-Friendly Benefits      &     -0.409***&     -0.414***&     -0.409***&      -0.420***&    -0.305***&     -0.379***&     -0.293***\\
                              &    (0.017)   &    (0.018)   &    (0.017)   &     (0.017)   &   (0.017)   &    (0.018)   &    (0.018)   \\
[1em]
Company Culture Ratings       &              &      0.034   &              &              &              &              &     -0.022   \\
                              &              &    (0.033)   &              &              &              &              &    (0.042)   \\
[1em]
Career Growth Ratings         &              &              &      0.001   &              &              &              &     -0.084** \\
                              &              &              &    (0.033)   &              &              &              &    (0.040)   \\
[1em]
Management Ratings            &              &              &              &      0.294***&              &              &      0.374***\\
                              &              &              &              &    (0.051)   &              &              &    (0.061)   \\
[1em]
Log Salary                    &              &              &              &              &     -0.692***&              &     -0.623***\\
                              &              &              &              &              &    (0.021)   &              &    (0.022)   \\
\hline
N Job Spells                  &     53479    &      53479   &      53479   &      53479   &      53479   &      53479   &      53479   \\
Industry Indicators           &          -   &          -   &          -   &          -   &          -   &        Yes   &        Yes   \\
\bottomrule[2pt]
\end{tabular}
}
}
    \endgroup

    \vspace{24pt}
    \caption*{\normalsize 
        This table reports coefficients and standard errors (in parentheses) from the proportional hazards model for the relationship between family-friendly benefits and separation hazards (discussed in \autoref{subsec:separation_survival}). 
        Each observation is a job spell that records tenure at separation in months since hire. 
        Covariates in the cox model include an indicator for family-friendly benefits, log starting salary, industry indicators, and 5-point likert scales for company ratings. 
        The sample consists of job spells with childbirths during tenure and non-missing data on company ratings through TeamBlind. 
        *, **, and *** denote statistical significance at 10\%, 5\%, and 1\% levels, respectively.
    }
\end{table}
\vspace*{\fill}
\clearpage

\restoregeometry


\end{document}